\documentclass[10pt,a4paper,superscriptaddress,aps,nofootinbib,showkeys,showpacs]{revtex4-2}

\usepackage{amsmath,amssymb,graphicx}
\usepackage[active]{srcltx}
\usepackage{graphicx,color}
\usepackage{changebar}
\usepackage{hyperref}
\usepackage{placeins}
\hypersetup{
	colorlinks=true,
	urlcolor= blue,
	citecolor=blue,
	linkcolor= blue,
	bookmarks=true,
	bookmarksopen=false,
}
\usepackage[T1]{fontenc}
\usepackage{ae}
\usepackage[utf8]{inputenc}
\usepackage{braket}
\usepackage{mathtools}
\usepackage{booktabs}
\usepackage{slashed}
\usepackage{empheq}
\usepackage{tikz}
\usepackage{multirow}
\usetikzlibrary{decorations.pathmorphing}
\usepackage{amsmath,amssymb,graphics,epsfig,subfigure}
\usetikzlibrary{quotes,angles}
\usepackage{etoolbox}
\usepackage{orcidlink}
\usepackage{soul}
\usepackage[normalem]{ulem}

\usetikzlibrary{arrows}
\usetikzlibrary{decorations.markings}

\begin{document}
	
	\date{\today}
	
	\title{Numerical study of non-relativistic quantum systems and small oscillations induced in a helically twisted geometry}
	
	\author{C. F. S. Pereira \orcidlink{0000-0001-6913-0223}}
	\email{carlosfisica32@gmail.com}
	\affiliation{Departamento de F\'isica e Qu\'imica, Universidade Federal do Esp\'irito Santo, Avenida Fernando Ferrari, 514, Goiabeiras, 29060-900, Vit\'oria, Esp\'irito Santo, Brazil.}
	
	\author{R. L. L. Vit\'oria \orcidlink{0000-0001-8802-3634}}
	\email{ricardovitoria@professor.uema.br/ricardo-luis91@hotmail.com}
	\affiliation{Programa de P\'os-Gradua\c c\~ao em Engenharia Aeroespacial, Universidade Estadual do Maranh\~ao,
		Cidade Universit\'aria Paulo VI, S\~ao Lu\'is, MA 65055-310, Brazil}
	\affiliation{Faculdade de F\'isica, Universidade Federal do Par\'a,
		Av. Augusto Corr\^ea, Guam\'a, Bel\'em, PA 66075-110, Brazil}
	
	\author{A. R. Soares \orcidlink{0000-0003-1871-2068}}
	\email{adriano.soares@ifma.edu.br}
	\affiliation{Grupo de Investiga\c{c}\~{a}o em Educa\c{c}\~{a}o Matem\'{a}tica, Instituto Federal de Educa\c{c}\~ao Ci\^encia e Tecnologia do Maranh\~ao,  Rua  Deputado Gast\~ao Vieira, 1000, CEP 65393-000 Buriticupu, MA, Brazil.}
	
	\author{B.B. Silva \orcidlink{0000-0001-7346-0129}}
	\email{breno.silva.80@edu.ufes.br}
	\affiliation{Departamento de F\'isica e Qu\'imica, Universidade Federal do Esp\'irito Santo, Avenida Fernando Ferrari, 514, Goiabeiras, 29060-900, Vit\'oria, Esp\'irito Santo, Brazil.}
	
	\author{H. Belich \orcidlink{0000-0002-8795-1735}}
	\email{humberto.belich@ufes.br}
	\affiliation{Departamento de F\'isica e Qu\'imica, Universidade Federal do Esp\'irito Santo, Avenida Fernando Ferrari, 514, Goiabeiras, 29060-900, Vit\'oria, Esp\'irito Santo, Brazil.}
	
	\author{Edilberto O. Silva\orcidlink{0000-0002-0297-5747}}
	\email{edilberto.silva@ufma.br}
	\affiliation{Departamento de F\'{\i}sica, Universidade Federal do Maranh\~{a}o, 65085-580 S\~{a}o Lu\'{\i}s, Maranh\~{a}o, Brazil}
	
	\begin{abstract}
		We investigate bound states of a non-relativistic scalar particle in a three-dimensional helically twisted (torsional) geometry, considering both the free case and the presence of external radial interactions. The dynamics is described by the Schr\"odinger equation on a curved spatial background and, when included, by minimal coupling to a magnetic vector potential incorporating an Aharonov--Bohm flux. After separation of variables, the problem reduces to a one-dimensional radial eigenvalue equation governed by an effective potential that combines torsion-induced Coulomb-like and centrifugal-like structures with magnetic/flux-dependent terms and optional model interactions. Because closed-form analytic solutions are not reliable over the parameter ranges required for systematic scans, we compute spectra and eigenfunctions numerically by formulating the radial equation as a self-adjoint Sturm--Liouville problem and solving it with a finite-difference discretization on a truncated radial domain, with explicit convergence control. We analyze four representative scenarios: (i) no external potential, (ii) Cornell-type confinement, (iii) Kratzer-type interaction, and (iv) the small-oscillation regime around the minimum of a Morse potential. We present systematic trends of the low-lying levels as functions of the torsion parameter, magnetic field, and azimuthal sector, and we show that geometric couplings alone can produce effective confinement even in the absence of an external interaction.
	\end{abstract}
		
	
	\maketitle
	
	\section{Introduction}\label{sec1}
	
	In recent decades, the study of topological defects has attracted significant interest across several areas of physics and mathematics \cite{I1}. Besides their role as solutions of nonlinear differential equations, such defects may act as domain walls delimiting distinct regions, often with observable consequences. Representative examples include kinks, vortices, and skyrmion-like configurations \cite{I2,I3,I4,I5}. In gravitation and cosmology, cosmic strings arise in the Abelian Higgs model and share structural similarities with vortex solutions in flat spacetime, connecting spontaneous symmetry breaking to macroscopic phenomena \cite{I1,I3,I5}.
	
	The investigation of quantum dynamics in nontrivial backgrounds has produced a large literature, including relativistic and non-relativistic hydrogen-like problems in spacetimes containing topological defects such as cosmic strings and global monopoles \cite{I6,I7}. Several extensions address magnetic fields, Aharonov--Bohm-type effects, extra dimensions, position-dependent masses, and related spectral problems \cite{I8,I9,I10,I11,I12,I13,I14,I15,I16,I17,I18,I19,I20,I21,I22,I23,I24,I25,I27,I28,I29,I30,I31}. In these contexts, curvature and torsion can induce effective interactions and modify quantum phases. In particular, torsion is naturally described in Einstein--Cartan-type geometries and also appears in condensed-matter analogs such as defects in elastic or crystalline media \cite{put,valan,naba}.
	
	A twisted helical geometry has been proposed as a tractable model of a torsion-like defect and explored for bound states, numerical classification of levels, and related physical applications \cite{1,2,3}. The present work focuses on a consistent numerical treatment of the bound-state problem in this background. Instead of relying on series truncations and special-function representations, we formulate the reduced radial equation as a self-adjoint Sturm--Liouville eigenvalue problem and compute the spectrum using finite-difference discretization with explicit convergence checks. This strategy is robust across parameter regimes and directly supports scans required for spectral plots.
	
	The paper is organized as follows. In Section~\ref{model} we introduce the helically twisted metric. In Section~\ref{LANDAU} we derive the reduced radial equation from the curved-space Schr\"odinger equation with minimal coupling to a magnetic vector potential and Aharonov--Bohm flux, and we present the numerical eigenvalue formulation and discretization scheme used throughout. Sections~\ref{sec4}--\ref{sec6} apply the same numerical machinery to external Cornell and Kratzer interactions and to the small-oscillation (quadratic-plus-linear) approximation of the Morse potential, respectively. Section~\ref{sec7} summarizes the main numerical findings and physical trends.
	
	\section{Geometric model: helically twisted space}\label{model}
	
	We consider an axially symmetric three-dimensional geometry characterized by a torsion-like helical twist that couples the angular coordinate $\phi$ to the longitudinal coordinate $z$. The spatial line element is
	\begin{equation}
		ds^2= dr^2 +r^2(1+\omega^2)d\phi^2 +2\omega r\,dz\,d\phi + dz^2,
		\label{eq:metric}
	\end{equation}
	where $\omega$ is a dimensionless parameter controlling the strength of the helical twist. The coordinates are $0<r<\infty$, $0<\phi\leq2\pi$, and $-\infty<z<\infty$. The metric tensor and its inverse are
	\begin{equation}\label{eq:gij}
		g_{ij }=\left(
		\begin{tabular}{ccc}
			$1$ & $0$ & $0$ \\
			$0$ & $r^2(1+\omega^2)$ & $\omega r$ \\
			$0$ & $\omega r$ & $1$
		\end{tabular}
		\right), \quad
		g^{ij }=\left(
		\begin{tabular}{ccc}
			$1$ & $0$ & $0$ \\
			$0$ & $\frac{1}{r^2}$ & $-\frac{\omega}{r}$ \\
			$0$ & $-\frac{\omega}{r}$ & $1+\omega^2$
		\end{tabular}
		\right),
	\end{equation}
	with determinant $g=\det(g_{ij})=r^2$. Further geometric and physical properties of this background can be found in Refs.~\cite{1,2,3}.
	
	\section{Reduced radial equation and numerical eigenvalue problem}\label{LANDAU}
	
	\subsection{Curved-space Schr\"odinger equation with minimal coupling}
	
	We start from the stationary Schr\"odinger equation for a scalar particle of effective mass $\mu$ in the curved spatial background,
	\begin{equation}\label{eq:Sch_general}
		\left(E-V(r)\right)\psi= -\frac{\hbar^2}{2\mu}\frac{1}{\sqrt{g}}D_k\!\left[\sqrt{g}\,g^{ki}D_i\right]\psi,
	\end{equation}
	where $g=\det(g_{ij})$ and $D_i=\partial_i-i e A_i$ implements minimal coupling to the electromagnetic potential. We adopt a uniform magnetic field along $z$ and include an Aharonov--Bohm flux through
	\begin{equation}\label{eq:Avec}
		A_\varphi= -\frac{B_0}{2}r^2 + \frac{\Phi_B}{2\pi}, \qquad  A_t=A_z=A_r=0,
	\end{equation}
	with constant $B_0$ and flux $\Phi_B$ \cite{6,7,I22}.
	
	Using the ansatz
	\begin{equation}\label{eq:sep}
		\psi(r,\phi,z)=e^{i m\phi}e^{i k z}\,\xi(r),
	\end{equation}
	with $m\in\mathbb{Z}$ and $k\in\mathbb{R}$, Eq.~\eqref{eq:Sch_general} reduces to
	\begin{equation}\label{eq:xi_eq}
		\xi''(r)+\frac{1}{r}\xi'(r)+\left[\frac{2\mu}{\hbar^2}\left(E-V(r)\right)-V_1(r)\right]\xi(r)=0,
	\end{equation}
	where
	\begin{align}\label{eq:V1}
		&V_1(r)=\;\frac{m^2}{r^2}+k^2(1+\omega^2)-\frac{2m\omega k}{r}
		+\frac{2\omega k e}{r}\left(-\frac{B_0 r^2}{2}+\frac{\Phi_B}{2\pi}\right)-\frac{2 m e}{r^2}\left(-\frac{B_0 r^2}{2}+\frac{\Phi_B}{2\pi}\right)
		+\frac{e^2}{r^2}\left(-\frac{B_0 r^2}{2}+\frac{\Phi_B}{2\pi}\right)^2.
	\end{align}
	
	Equation (\ref{eq:V1}) represents a potential energy term, which carries the geometric and electromagnetic information of the system. It is important to note that, if we set $e\rightarrow0$, the potential energy $V_1$ falls into $V_1=\frac{m^2}{r^2}+k^2(1+\omega^2)\notag-\frac{2m\omega k}{r}$, that is, a potential energy of a purely geometric nature. This characteristic is provided by the presence of the parameter $\omega$ in the potential energy. Furthermore, we can note that this interaction contains a Coulomb-type potential plus a centrifugal-type potential term, thus forming a geometric Kratzer-Fues-type potential, $V_{KF}$, defined in form $V_{KF}=\frac{m^2}{r^2}-\frac{2m\omega k}{r}$. In this particular case, taking $k=0$, we can note that the potential $V_1$ is reduced even further, becoming a centrifugal-type potential and, consequently, the geometric effects are inhibited; the quantum particle is not influenced by the torsion of spacetime.
	
	It is convenient to eliminate the first-derivative term by $\xi(r)=f(r)/\sqrt{r}$. Then Eq.~\eqref{eq:xi_eq} becomes a one-dimensional Schr\"odinger-like equation,
	\begin{equation}\label{eq:f_eq_basic}
		f''(r)+\left[\frac{2\mu}{\hbar^2}\left(E-V_{\mathrm{eff}}(r)\right)\right]f(r)=0,
	\end{equation}
	with
	\begin{equation}
		V_{\mathrm{eff}}(r)=V(r)+\frac{\hbar^2}{2\mu}\left(V_1(r)-\frac{1}{4r^2}\right).
		\label{eq:Veff-general}
	\end{equation}
	This redefinition is advantageous because it casts the radial problem into a standard self-adjoint one-dimensional form, where the spectral parameter $E$ enters linearly and all geometric and electromagnetic effects are encoded in a single effective potential. The extra contribution $-\,\hbar^2/(8\mu r^2)$ originates solely from the removal of the first derivative and should be regarded as a universal short-distance term of the reduced description, independent of the external model interaction $V(r)$. In this form, the bound-state problem is naturally treated as an eigenvalue equation on a truncated radial interval with appropriate boundary conditions, which is precisely the setup adopted in our finite-difference Sturm--Liouville implementation in the next subsection.
	
	\subsection{Sturm--Liouville formulation}
	
	For numerical purposes, we recast Eq.~\eqref{eq:f_eq_basic} as a self-adjoint eigenvalue problem. Define
	\begin{equation}\label{eq:SL}
		\hat{\mathcal{A}} f(r)=\lambda f(r),\qquad
		\hat{\mathcal{A}}=-\frac{d^2}{dr^2}+U(r),
	\end{equation}
	with
	\begin{equation}\label{eq:U_def}
		U(r)=\frac{2\mu}{\hbar^2}V_{\mathrm{eff}}(r),\qquad \lambda=\frac{2\mu}{\hbar^2}E.
	\end{equation}
	Thus, once the lowest eigenpairs $(\lambda_n,f_n)$ are obtained, the physical energies follow from $E_n=(\hbar^2/2\mu)\lambda_n$ and the full wavefunction is reconstructed using Eq.~\eqref{eq:sep} and $\xi_n(r)=f_n(r)/\sqrt{r}$.
	
	For later use in the discussion of eigenfunctions, we also define the (radial) probability density associated with the reduced function $f(r)$,
	\begin{equation}\label{eq:rho_def}
		\rho(r)\equiv |f(r)|^2,
	\end{equation}
	together with the normalization condition on the truncated domain,
	\begin{equation}\label{eq:norm_f}
		\int_{r_{\min}}^{r_{\max}} |f(r)|^2\,dr=1.
	\end{equation}
	When comparing different parameters, all reported densities are computed from eigenfunctions normalized according to Eq.~\eqref{eq:norm_f}.
	
	\subsection{Finite-difference discretization and boundary conditions}
	
	We solve Eq.~\eqref{eq:SL} on a truncated domain $r\in[r_{\min},r_{\max}]$ with a uniform grid
	\begin{equation}
		r_i=r_{\min}+i\Delta r,\;\; i=0,1,\ldots,N,\;\; \Delta r=\frac{r_{\max}-r_{\min}}{N}.
	\end{equation}
	The second derivative is approximated by the standard three-point stencil
	\begin{equation}
		\left.\frac{d^2 f}{dr^2}\right|_{r=r_i}\approx \frac{f_{i+1}-2f_i+f_{i-1}}{\Delta r^2},
	\end{equation}
	which yields a real symmetric tridiagonal matrix representation of $\hat{\mathcal{A}}$ in Eq.~\eqref{eq:SL}. Throughout this work we impose Dirichlet boundary conditions,
	\begin{equation}\label{eq:BC}
		f(r_{\min})=0,\qquad f(r_{\max})=0.
	\end{equation}
	The inner cutoff $r_{\min}>0$ regularizes the $1/r$ and $1/r^2$ terms numerically; convergence is verified by decreasing $r_{\min}$ and refining $\Delta r$. The outer boundary $r_{\max}$ is chosen large enough that the low-lying bound states are exponentially small at $r_{\max}$, ensuring that Eq.~\eqref{eq:BC} does not distort the spectrum.
	
	In all parameter scans, numerical reliability is ensured by:
	(i) grid refinement tests ($N\to 2N$ at fixed $r_{\min},r_{\max}$),
	(ii) domain enlargement tests ($r_{\max}\to r_{\max}+\Delta$ at fixed $\Delta r$),
	and (iii) stability of the first few eigenvalues within a prescribed tolerance. The tridiagonal structure of the discretized operator allows efficient computation of the lowest eigenvalues using dedicated symmetric-tridiagonal eigensolvers.
	
	\subsection{Model I: no external potential}
	
	We start with the simplest scenario, in which the particle dynamics is governed solely by the geometry of the background and by the uniform magnetic field, with no additional external radial interaction. In other words, by setting $V(r)=0$ in Eq.~\eqref{eq:Veff-general}, we obtain an effective potential that encodes only centrifugal, torsion-induced, and magnetic contributions, namely
	\begin{equation}\label{eq:Veff-I}
		V_{\mathrm{eff}}^{(\mathrm{I})}(r)
		=\frac{\hbar^{2}}{2\mu}\,%
		\left[
		\frac{m^{2}-\tfrac{1}{4}}{r^{2}}
		+k^{2}(1+\omega^{2})
		-\frac{2m\omega k}{r}
		+\frac{2\omega k e}{r}\,A_{\varphi}(r)
		-\frac{2me}{r^{2}}\,A_{\varphi}(r)
		+\frac{e^{2}}{r^{2}}\,A_{\varphi}^{2}(r)
		\right].
	\end{equation}
	This potential exhibits a clear physical interplay among its radial terms: the first one corresponds to the centrifugal barrier ($\propto 1/r^2$), the second one represents a torsion-induced Coulomb-like contribution ($\propto -1/r$, whose sign depends on $m\omega k$), while the last term provides the usual magnetic (harmonic) confinement ($\propto r^2$). In addition, the linear contribution in $r$ (proportional to $-\omega k eB_0$) tilts the profile, shifting the position of the minimum.
	
\begin{figure}[!ht]
		\centering
		\includegraphics[width=0.48\linewidth]{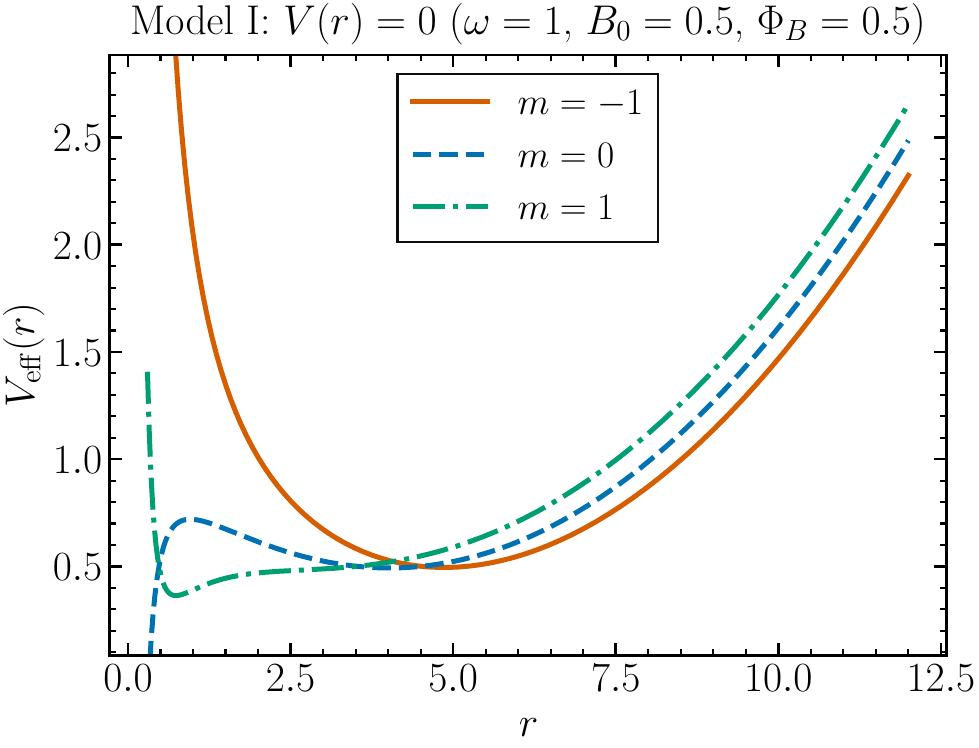}
		\caption{Effective potential $V_{\mathrm{eff}}^{(\mathrm{I})}(r)$
			[Eq.~\eqref{eq:Veff-I}] for Model~I ($V(r)=0$) in the helically twisted geometry
			in the presence of a uniform magnetic field and an Aharonov--Bohm flux.
			The azimuthal vector potential is kept throughout,
			$A_{\varphi}(r)=-(B_0/2)\,r^{2}+\Phi_{B}/(2\pi)$.
			The curves correspond to $m=-1,0,1$ (as labeled in the plot).
			At large $r$, the magnetic contribution contained in $A_\varphi$ produces an effective parabolic wall,
			ensuring confinement in all sectors.
			At short distances, the dominant $1/r^{2}$ structure (including the $-1/(4r^2)$ term and the
			$A_\varphi$--dependent pieces) competes with the torsion Coulomb--like term $\propto -2m\omega k/r$,
			leading to an $m$--dependent shift of the well depth and its equilibrium position.
			Parameters: $\hbar=\mu=e=k=1$, $\omega=1$, $B_0=0.5$, and $\Phi_B=0.5$ (dimensionless units).}
		\label{fig:Veff-I}
	\end{figure}

	For $B_0\neq 0$ (and keeping the vector potential sector active), the joint action of the short--distance
	$1/r^2$ barrier, the torsion--induced Coulomb--like contribution $\propto -2m\omega k/r$, and the magnetic
	confinement generated by $A_\varphi(r)$ yields a well--defined potential well (stable minimum) in each
	azimuthal sector, as illustrated in Fig.~\ref{fig:Veff-I}.
	
	Figure~\ref{fig:Veff-I} shows that, even for $V(r)=0$, the geometry plus the electromagnetic sector generate
	a confining effective potential with a clear minimum and a rising wall at large $r$.
	The large--$r$ behavior is governed by the magnetic confinement encoded in $A_\varphi(r)$, which yields an
	effective contribution growing as $\propto r^{2}$.
	The splitting among the $m=-1,0,1$ curves is produced by the combined $m$--dependent centrifugal/torsional
	terms and by the couplings to $A_\varphi$ (linear in $A_\varphi$ and quadratic in $A_\varphi$),
	so changing $m$ mainly shifts the location and depth of the minimum while preserving the overall confining profile.
	In particular, the torsion term $\propto -2m\omega k/r$ is attractive for $m>0$ and repulsive for $m<0$,
	which contributes to deepening/shallower wells in the $m=1$/$m=-1$ sectors, respectively, on top of the
	universal magnetic confinement at large radii.

	\begin{figure*}[tbhp]
		\centering
		\includegraphics[width=0.98\linewidth]{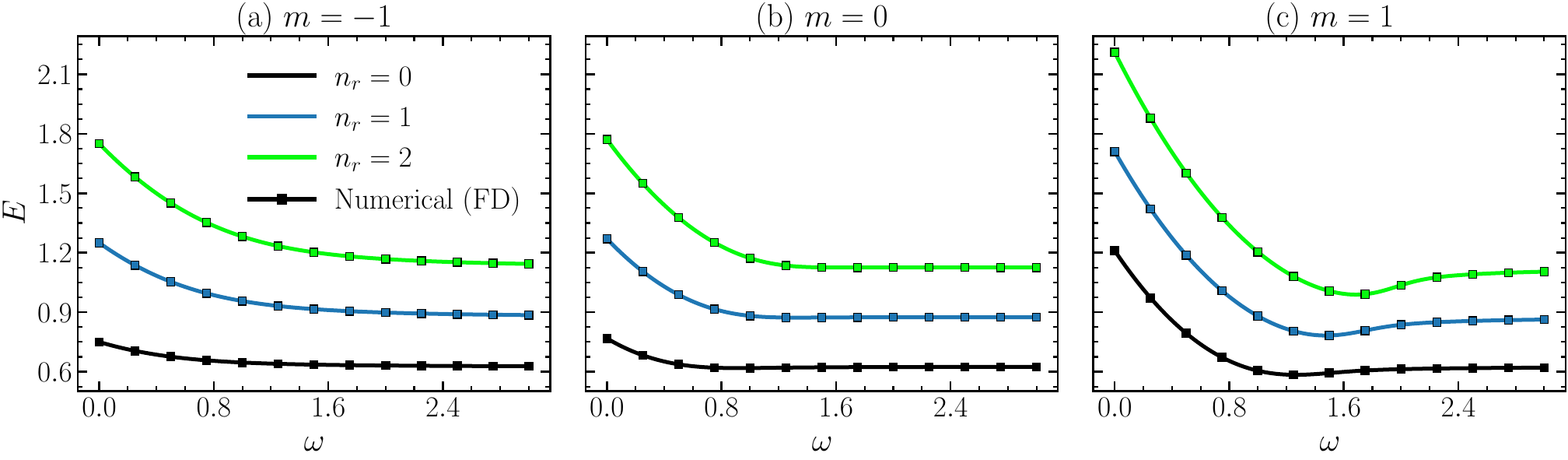}
		\caption{
			Numerical (finite-difference) energy spectrum $E$ as a function of the torsion parameter $\omega$
			in the helically twisted geometry for Model~I ($V(r)=0$).
			The energies are obtained by solving the radial Sturm--Liouville eigenvalue problem~\eqref{eq:SL}
			with Dirichlet boundary conditions~\eqref{eq:BC}, using the full geometric/gauge contribution
			encoded in $V_1(r)$ [Eq.~\eqref{eq:V1}] through the effective potential $V_{\mathrm{eff}}(r)$
			[Eq.~\eqref{eq:Veff-general}].
			The three panels correspond to fixed azimuthal quantum numbers:
			(a) $m=-1$, (b) $m=0$, and (c) $m=1$.
			In each panel we display the lowest three radial levels, $n_r=0,1,2$ (color-coded),
			computed from the tridiagonal finite-difference discretization described in Sec.~\ref{LANDAU}.
			Parameters: $\hbar=\mu=e=k=1$, $B_0=0.5$, and $\Phi_B=0.5$ (dimensionless units).
		}
		\label{fig:E_vs_omega_ModelI}
	\end{figure*}
	
	Figure~\ref{fig:E_vs_omega_ModelI} summarizes the bound-state spectrum produced purely by the
	geometric (torsional) coupling in the reduced radial problem.
	For each fixed $m$, the eigenenergies decrease monotonically as $\omega$ increases,
	reflecting the strengthening of the effective confinement induced by the helical distortion.
	For $m=\pm1$ [Figs.~\ref{fig:E_vs_omega_ModelI}(a) and \ref{fig:E_vs_omega_ModelI}(c)], the spectra display the same qualitative evolution with $\omega$ but are, in general, shifted relative to each other because the universal contribution $V_1(r)$ contains terms linear in $m$ through its couplings to $A_{\varphi}(r)$ when $(B_0,\Phi_B)\neq(0,0)$. An exact $m\to -m$ symmetry is recovered only in the gauge-decoupled limit (e.g., $e=0$) or when the vector-potential sector is switched off.
	In contrast, the $m=0$ sector [Fig. \ref{fig:E_vs_omega_ModelI}(b)] exhibits a near-degeneracy of the lowest levels in the
	small-$\omega$ region, and the separation between radial excitations becomes more pronounced as
	$\omega$ grows, indicating that torsion progressively amplifies the radial quantization scale.
	Overall, the numerical curves provide the reference dataset for all subsequent analyses, since no
	analytical truncation condition is imposed in this numerical-only approach.
	
	\begin{figure}[tbhp]
		\centering
		\includegraphics[width=0.45\linewidth]{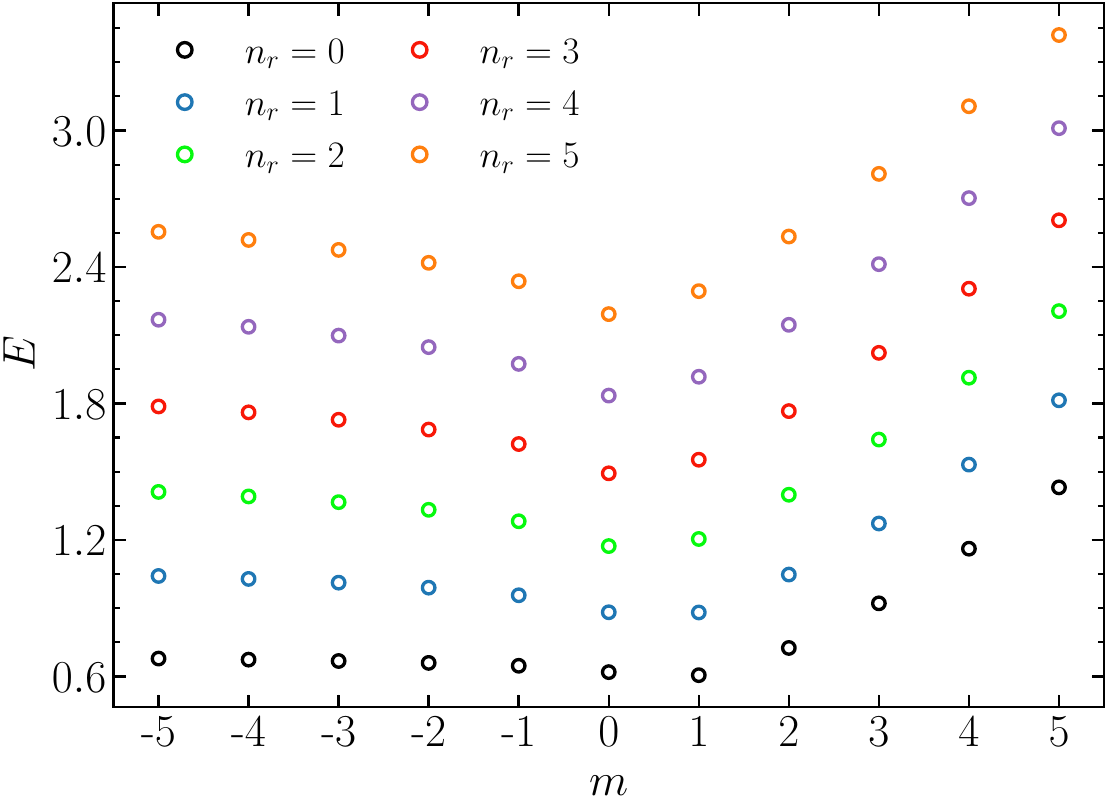}
		\caption{
			Discrete numerical spectrum $E$ as a function of the azimuthal quantum number $m$ for Model~I ($V(r)=0$) in the helically twisted geometry.
			For each integer $m\in[-5,5]$ we display the lowest six radial levels $n_r=0,1,2,3,4,5$ (color-coded markers).
			The eigenvalues are obtained by solving the radial Sturm--Liouville problem~\eqref{eq:SL} with Dirichlet boundary conditions~\eqref{eq:BC}, using a tridiagonal finite-difference discretization and direct diagonalization.
			Parameters: $\hbar=\mu=e=k=1$, $\omega=1$, $B_0=0.5$, and $\Phi_B=0.5$.
			The radial grid is uniform in $r\in[r_{\min},r_{\max}]$.
		}
		\label{fig:Evsm_ModelI}
	\end{figure}
	
	Figure~\ref{fig:Evsm_ModelI} shows the discrete set of bound-state energies obtained numerically as the azimuthal quantum number $m$ is varied at fixed torsion $\omega=1$.
	For each $m$, the reduced radial equation yields a ladder of eigenvalues labeled by $n_r$, and we plot the six lowest modes.
	Since the computation is performed directly at the level of the Sturm--Liouville eigenproblem, the spectrum is obtained without invoking any closed-form quantization or series-truncation conditions.
	The dependence on $m$ is governed by the interplay among the centrifugal-like contribution, the torsion-induced Coulomb-like structure, and the magnetic/flux background encoded in the universal geometric term $V_1(r)$ [Eq.~\eqref{eq:V1}] entering $V_{\mathrm{eff}}(r)$ [Eq.~\eqref{eq:Veff-general}].
	In particular, because the full $V_1(r)$ is retained (with $B_0\neq0$ and $\Phi_B\neq0$), the spectrum is not expected to be symmetric under $m\to -m$ in general, and the $m$-resolved splittings provide a direct diagnostic of how the helically twisted geometry and gauge background reshape the low-lying levels.
	Overall, the plot provides a compact numerical characterization of the $m$-resolved structure of the bound-state spectrum in Model~I for fixed $(\omega,B_0,\Phi_B)$.
	
	\begin{figure*}[tbhp]
		\centering
		\includegraphics[width=0.98\linewidth]{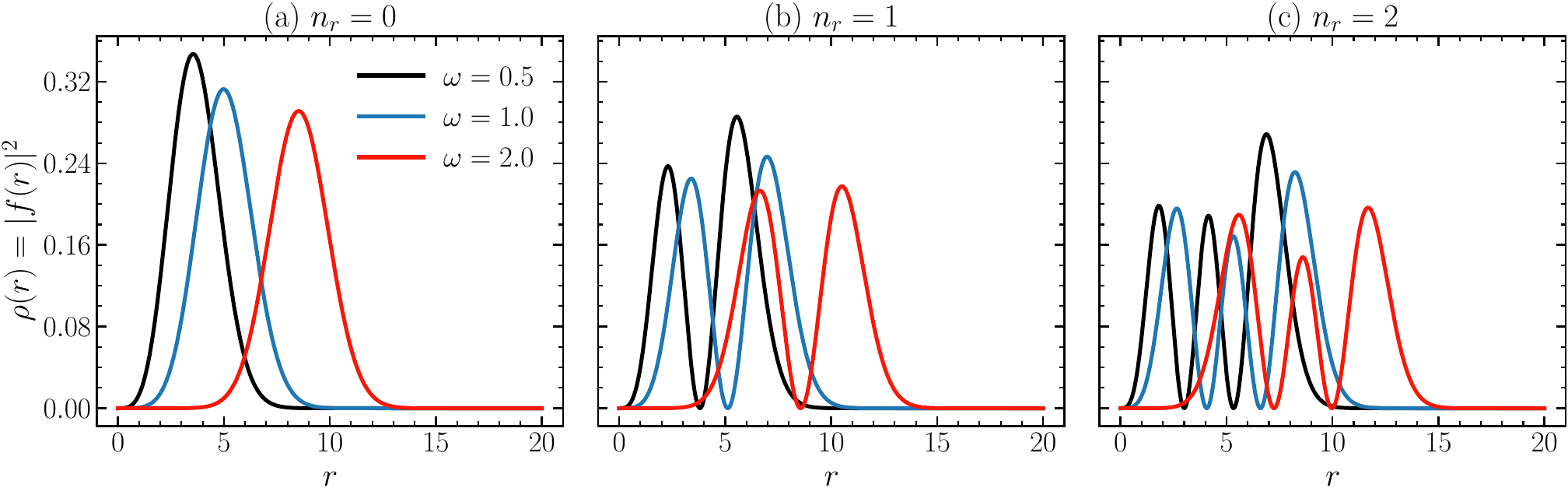}
		\caption{
			Numerical probability densities $\rho(r)=|f(r)|^2$ (reduced radial problem) for Model~I ($V(r)=0$)
			in the helically twisted geometry, obtained from the finite-difference solution of the
			Sturm--Liouville problem~\eqref{eq:SL} with Dirichlet boundary conditions~\eqref{eq:BC}.
			We fix the azimuthal sector to $m=-1$ and display three representative torsion values:
			$\omega=0.5$ (black), $\omega=1.0$ (blue), and $\omega=2.0$ (red).
			Panels (a)--(c) correspond to the lowest radial modes $n_r=0$, $n_r=1$, and $n_r=2$, respectively.
			The eigenfunctions are normalized according to Eq.~\eqref{eq:norm_f}.
			Parameters: $\hbar=\mu=e=k=1$, $B_0=0.5$, and $\Phi_B=0.5$.
		}
		\label{fig:rho_vs_r_ModelI_mminus1}
	\end{figure*}
	Figure~\ref{fig:rho_vs_r_ModelI_mminus1} shows how the torsion parameter $\omega$ reshapes the spatial
	localization of the reduced radial states in Model~I when the full geometric/gauge contribution
	$V_1(r)$ is retained (including the uniform magnetic field and the Aharonov--Bohm flux).
	As $\omega$ increases, the competition between the torsion-induced Coulomb-like structure, the
	centrifugal-like term, and the magnetic (parabolic) confinement modifies both the position and the
	width of the dominant probability peaks.
	In the ground state [Fig.~\ref{fig:rho_vs_r_ModelI_mminus1}(a)], the probability density becomes more
	concentrated and its maximum shifts, indicating enhanced localization in the region where the
	effective well is deepest for the chosen sector $m=-1$.
	For the excited states [Figs.~\ref{fig:rho_vs_r_ModelI_mminus1}(b) and
	\ref{fig:rho_vs_r_ModelI_mminus1}(c)], the expected nodal structure is recovered (one node for $n_r=1$
	and two nodes for $n_r=2$), while increasing $\omega$ compresses the oscillatory pattern and reduces
	the radial extent of the outer lobes.
	Overall, these profiles provide a direct numerical characterization of how torsion, in combination
	with the magnetic and flux backgrounds, controls the degree of radial localization across different
	$n_r$ sectors.

	\section{Model II: Cornell-type external potential}\label{sec4}
	
	We now move beyond the purely geometric/gauge-driven dynamics of Model~I by introducing an additional, standard confining interaction in the radial direction. Concretely, we supplement the effective interaction encoded in $V_{\mathrm{eff}}(r)$ [Eq.~\eqref{eq:Veff-general}] with a Cornell-type external potential,
	\begin{equation}\label{eq:Cornell}
		V(r)=\frac{a}{r}+b\,r,
	\end{equation}
	where $a>0$ and $b>0$ control, respectively, the short-range Coulomb sector and the long-range linear confinement.
	
	The purpose of this model is to assess how an \emph{explicit} Cornell interaction reshapes the low-lying spectrum when superposed on the intrinsic Coulomb-like structure already induced by torsion (and on the magnetic contributions). Importantly, the numerical strategy remains exactly the same as in Model~I: for each parameter set, we build $U(r)$ from Eqs.~\eqref{eq:Veff-general} and \eqref{eq:U_def}, discretize the Sturm--Liouville problem in Eq.~\eqref{eq:SL} using finite differences, and obtain the lowest eigenvalues subject to the boundary conditions~\eqref{eq:BC}. Parameter scans are then performed by repeating the same tridiagonal diagonalization while varying $(a,b)$ and/or $\omega$.
	
	With the Cornell interaction~\eqref{eq:Cornell}, the effective potential reads
	\begin{align}\label{eq:Veff-II}
		V_{\mathrm{eff}}^{(\mathrm{II})}(r)
		&=\frac{a}{r}+b\,r
		+\frac{\hbar^{2}}{2\mu}\,%
		\left[
		\frac{m^{2}-\tfrac{1}{4}}{r^{2}}
		+k^{2}(1+\omega^{2})
		-\frac{2m\omega k}{r}
		+\frac{2\omega k e}{r}\,A_{\varphi}(r)
		-\frac{2me}{r^{2}}\,A_{\varphi}(r)
		+\frac{e^{2}}{r^{2}}\,A_{\varphi}^{2}(r)
		\right]\,.
	\end{align}
	This expression makes the hierarchy of contributions transparent. At short distances, the $1/r^2$ centrifugal barrier competes with the \emph{combined} Coulomb sector, formed by the external term $a/r$ and the torsion-induced Coulomb-like piece $\propto -1/r$. At larger distances, the linear Cornell contribution $b\,r$ effectively reinforces the long-range confinement, adding to the magnetic $r^2$ term (and to the linear-in-$r$ tilt already present when $B_0\neq 0$), thereby deepening and shifting the potential well.
	
	\begin{figure}[!ht]
		\centering
		\includegraphics[width=0.48\linewidth]{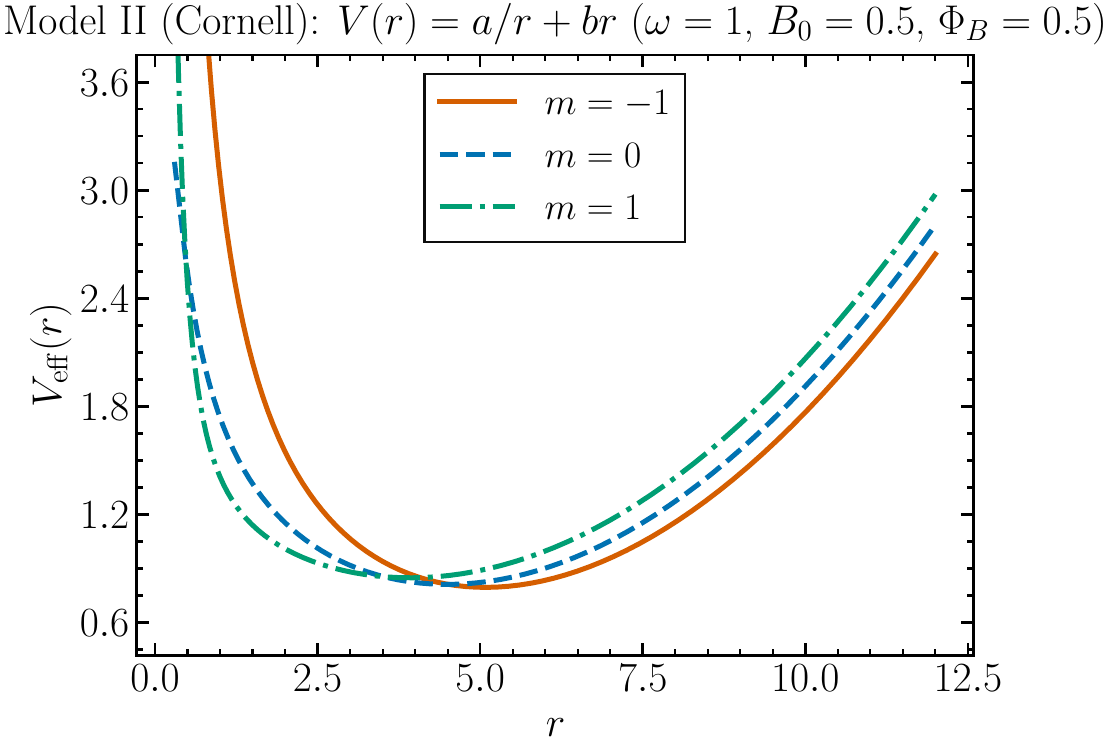}
		\caption{Effective potential $V_{\mathrm{eff}}^{(\mathrm{II})}(r)$
			[Eq.~\eqref{eq:Veff-II}] for Model~II (Cornell interaction, $V(r)=a/r+b\,r$)
			in the helically twisted geometry in the presence of a uniform magnetic field and an Aharonov--Bohm flux.
			The azimuthal vector potential is kept throughout,
			$A_{\varphi}(r)=-(B_0/2)\,r^{2}+\Phi_{B}/(2\pi)$.
			For $a>0$, the Coulomb term $a/r$ is \emph{repulsive} and increases the inner wall at short distances,
			competing with the torsion--induced Coulomb--like term $\propto -2m\omega k/r$, which is attractive for $m>0$
			and repulsive for $m<0$.
			At large radii, the linear confinement $b\,r$ adds to the magnetic parabolic wall (encoded in $A_\varphi$),
			leading to a steeper overall rise of $V_{\mathrm{eff}}^{(\mathrm{II})}(r)$.
			Parameters: $a=1.0$, $b=0.02$, $\hbar=\mu=e=k=1$, $\omega=1$, $B_0=0.5$, and $\Phi_B=0.5$
			(dimensionless units).}
		\label{fig:Veff-II}
	\end{figure}
	
	Figure~\ref{fig:Veff-II} displays $V_{\mathrm{eff}}^{(\mathrm{II})}(r)$ for $m=-1,0,1$.
	Compared with Model~I (Fig.~\ref{fig:Veff-I}), the Cornell interaction produces two robust changes.
	First, the repulsive Coulomb term $a/r$ strengthens the short--distance barrier for all sectors, shifting the
	inner turning point to larger radii; for $m=1$ this partially counterbalances the torsion--induced attraction
	$\propto -2m\omega k/r$, whereas for $m=-1$ it reinforces the overall repulsion.
	Second, the linear term $b\,r$ increases the outer slope and, together with the magnetic confinement generated
	by $A_\varphi(r)$, yields a faster growth of the effective potential at large $r$.
	As a consequence, the potential wells remain present but become typically narrower, and their depth/position
	exhibit an $m$--dependent displacement controlled by the combined torsion and electromagnetic couplings.
	The $m$--splitting is therefore preserved: the $m=-1$ sector is comparatively higher (shallower well), while
	the $m=1$ sector tends to lie lower (deeper well) over an intermediate radial range. 
	
	This potential is relevant in several contexts, including confinement modeling and related spectral problems \cite{I21,I25,4,24,25,26}. In the numerical framework, Eq.~\eqref{eq:Cornell} is inserted directly into $V_{\mathrm{eff}}(r)$ via Eq.~\eqref{eq:Veff-general}, and the spectrum is obtained by solving the same Sturm--Liouville problem~\eqref{eq:SL}.
	
	For reference, the reduced radial equation retains the general structure of Eq.~\eqref{eq:f_eq_basic}. In practice, we compute
	\begin{equation}
		U(r)=\frac{2\mu}{\hbar^2}\left[V(r)+\frac{\hbar^2}{2\mu}\left(V_1(r)-\frac{1}{4r^2}\right)\right]
	\end{equation}
	with $V(r)$ given by Eq.~\eqref{eq:Cornell}, and then diagonalize the resulting tridiagonal matrix.
	
	\begin{figure}[tbhp]
		\centering
		\includegraphics[width=0.45\linewidth]{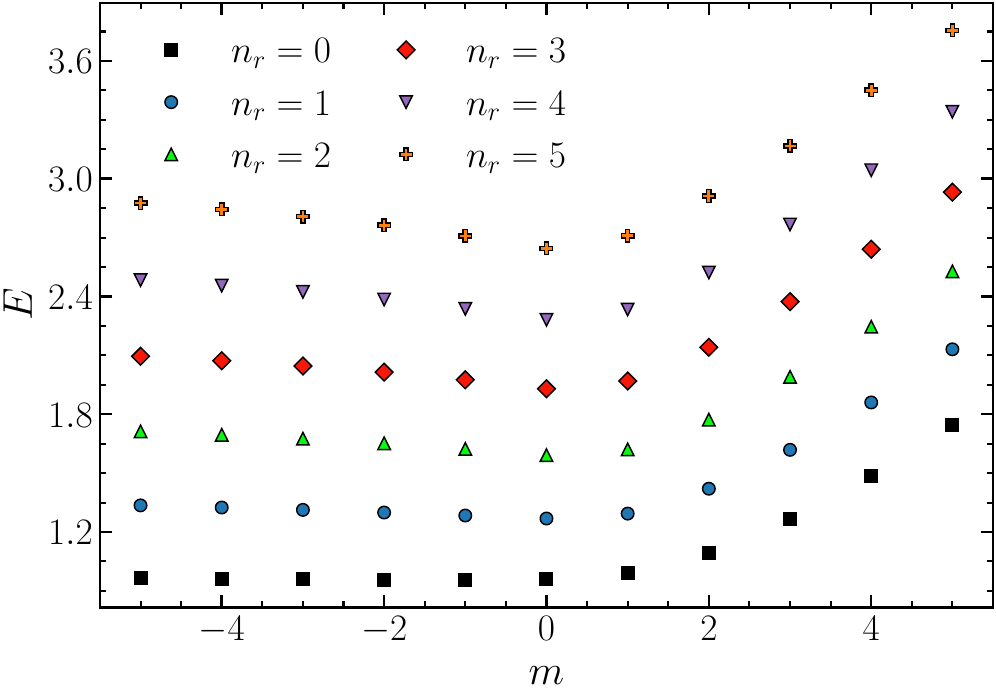}
		\caption{
			Bound-state energies $E$ as a function of the azimuthal quantum number $m$ for Model~II (Cornell confinement) in the helically twisted geometry.
			The external interaction is $V(r)=a/r+b\,r$, while the geometric/gauge background is fully retained through the universal term $V_1(r)$ entering $V_{\mathrm{eff}}(r)$ [Eqs.~\eqref{eq:V1} and \eqref{eq:Veff-general}], with a uniform magnetic field and Aharonov--Bohm flux.
			Discrete markers show the lowest six radial levels $n_r=0,1,2,3,4,5$ for each integer $m\in[-5,5]$.
			Parameters: $\hbar=\mu=e=k=1$, $\omega=1$, $a=1$, $b=0.02$, $B_0=0.5$, and $\Phi_B=0.5$.
		}
		\label{fig:Evsm_Cornell}
	\end{figure}
	
	Figure~\ref{fig:Evsm_Cornell} presents the $m$-resolved low-lying spectrum for the Cornell model, where the confinement produced by $V(r)=a/r+b\,r$ acts on top of the helically induced and gauge-induced structures encoded in $V_1(r)$.
	For each fixed $m$, the radial quantization yields an ordered ladder of levels labeled by $n_r$.
	In contrast to the $B_0=\Phi_B=0$ reduced setting, here the presence of the vector potential breaks, in general, the symmetry under $m\to -m$ because $V_1(r)$ contains terms linear in $m$ through its coupling to $A_\varphi(r)$.
	Consequently, the splittings between positive and negative $m$ sectors provide a direct measure of the interplay between the Cornell confinement and the magnetic/flux background.
	As $|m|$ increases, the combined short-distance structure (centrifugal-like core together with the Coulomb contributions from both torsion and Cornell terms) and the long-distance confinement (linear Cornell term plus magnetic parabolic growth) reshapes the level spacing and shifts the overall ladder, with higher radial excitations typically exhibiting a smoother dependence on $m$.
	Overall, the figure summarizes how the Cornell parameters $(a,b)$ and the torsion/gauge sector $(\omega,B_0,\Phi_B)$ cooperate to produce a tunable, $m$-dependent bound-state spectrum.

	\begin{figure*}[tbhp]
		\centering
		\includegraphics[width=\linewidth]{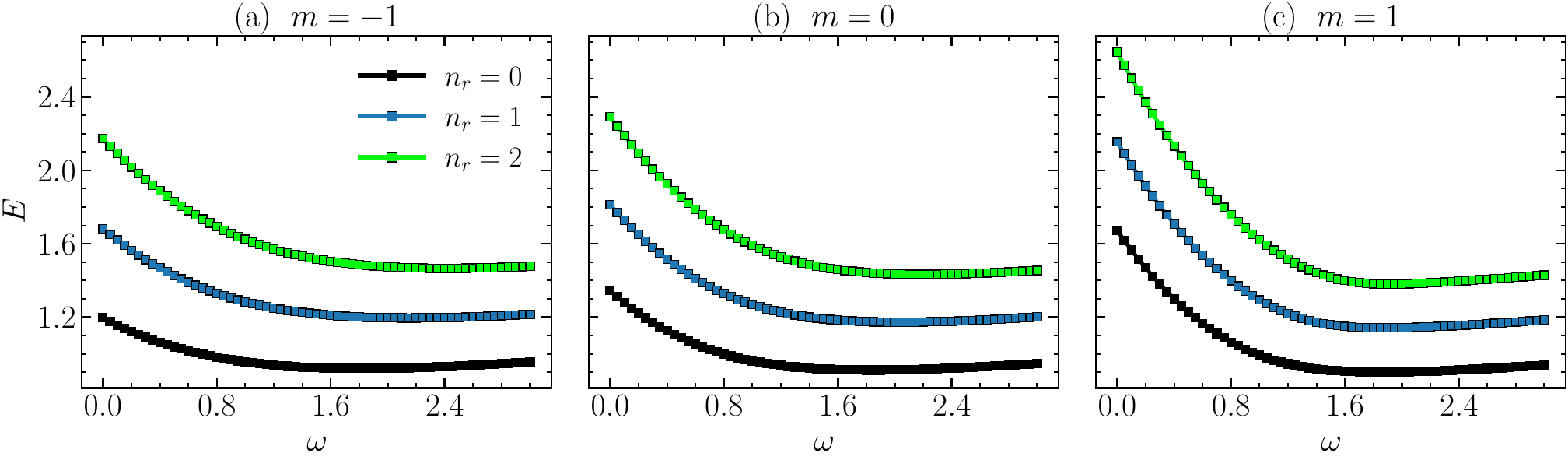}
		\caption{
			Bound-state energy spectrum $E$ as a function of the torsion parameter $\omega$ for the Cornell model
			$V(r)=a/r+b\,r$ [Eq.~\eqref{eq:Cornell}] in the helically twisted geometry, including the universal gauge contribution $V_1(r)$
			through the azimuthal vector potential $A_\varphi(r)$ (uniform field $B_0$ and Aharonov--Bohm flux $\Phi_B$).
			The panels correspond to (a) $m=-1$, (b) $m=0$, and (c) $m=1$.
			In each panel, the three lowest radial levels $n_r=0,1,2$ are displayed.
			The reduced Sturm--Liouville problem \eqref{eq:SL}--\eqref{eq:BC} is built from $V_{\mathrm{eff}}(r)$ in Eq.~\eqref{eq:Veff-general},
			where $V(r)$ is the Cornell interaction and $V_1(r)$ is given by Eq.~\eqref{eq:V1}.
			Parameters: $\hbar=\mu=e=k=1$, $a=1.0$, $b=0.02$, $B_0=0.5$, and $\Phi_B=0.5$ (dimensionless units).
		}
		\label{fig:E_vs_omega_Cornell}
	\end{figure*}
	
	Figure~\ref{fig:E_vs_omega_Cornell} shows how torsion reshapes the Cornell spectrum once the gauge sector is consistently included via
	$A_\varphi(r)$ (i.e., with $B_0\neq 0$ and $\Phi_B\neq 0$).
	As $\omega$ increases, the low-lying levels shift and the radial quantization scale changes, reflecting the strengthening of the effective
	confinement in $V_{\mathrm{eff}}(r)$ produced by the combined action of the Cornell interaction and the $\omega$-dependent terms inside $V_1(r)$.
	For each fixed $m$, the ordering $E_{0m}<E_{1m}<E_{2m}$ is preserved across the scanned interval, while the separation between successive radial
	excitations typically grows as $\omega$ becomes larger.
	Because the full $V_1(r)$ contains terms linear in $m$ through the couplings to $A_\varphi(r)$, the spectra in the $m=\pm1$ sectors are no longer
	constrained to be symmetric under $m\!\to\!-m$ when $(B_0,\Phi_B)\neq(0,0)$; consequently, panels (a) and (c) may exhibit systematically distinct
	energy shifts at the same $\omega$.
	Overall, the figure provides the $\omega$-resolved, $m$-resolved reference spectrum for the Cornell model in the helically twisted geometry with the
	gauge sector switched on.

	\begin{figure*}[tbhp]
		\centering
		\includegraphics[width=\linewidth]{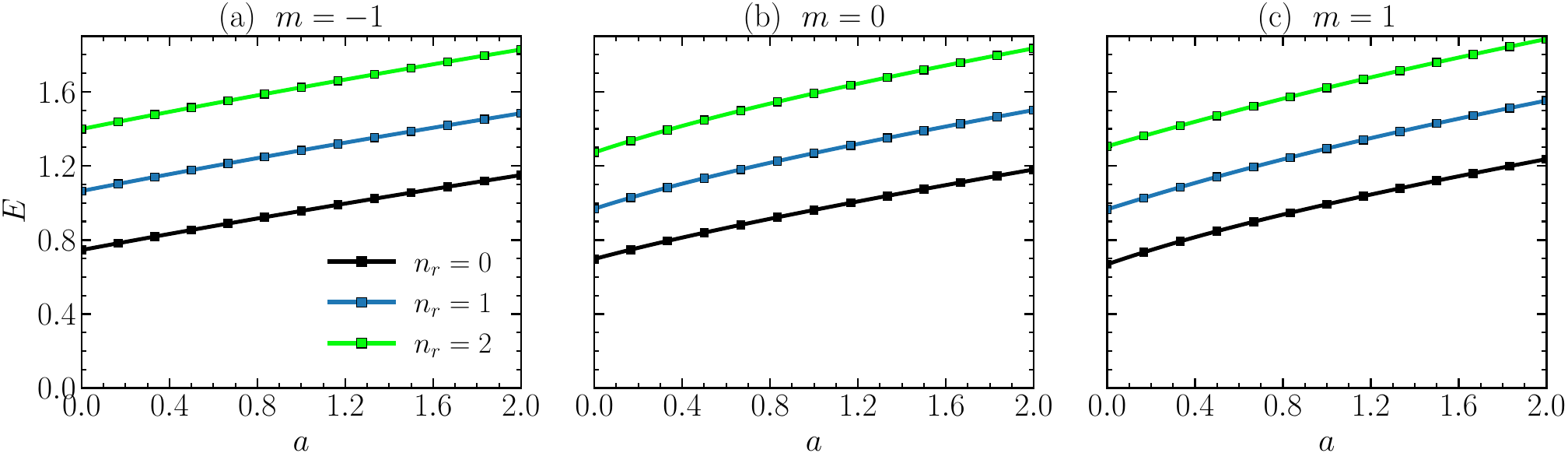}
		\caption{%
			Bound-state energies $E$ as a function of the Cornell parameter $a$ in the external interaction
			$V_{\mathrm{C}}(r)=a/r+b\,r$ [Eq.~\eqref{eq:Cornell}] for the helically twisted geometry (Model~II).
			The three panels correspond to fixed azimuthal sectors:
			(a) $m=-1$, (b) $m=0$, and (c) $m=1$.
			In each panel we display the three lowest radial levels $n_r=0,1,2$.
			The gauge sector is kept on ($B_{0}=0.5$ and $\Phi_{B}=0.5$).
			The remaining parameters are fixed to $\hbar=\mu=e=k=1$, $\omega=1$, and $b=0.02$ (dimensionless units).%
		}
		\label{fig:E_vs_a_Cornell}
	\end{figure*}
		
	Figure~\ref{fig:E_vs_a_Cornell} shows how the low-lying spectrum responds to the Coulomb-like Cornell coupling $a$,
	with all other parameters held fixed, including the torsion parameter $\omega$, the linear Cornell strength $b$,
	and the gauge parameters $(B_{0},\Phi_{B})$ entering $V_{1}(r)$.
	For all three azimuthal sectors, the eigenenergies decrease as $a$ increases, consistent with the fact that strengthening the attractive $a/r$ term deepens the effective radial well.
	For each fixed $m$, the radial ordering is preserved throughout the scanned interval,
	\begin{equation*}
		E_{0m}<E_{1m}<E_{2m}.
	\end{equation*}
	Comparing panels (a)--(c), the different vertical offsets and slopes reflect the explicit $m$-dependence of the universal contribution $V_{1}(r)$, which contains both centrifugal-like pieces and linear-in-$m$ couplings induced by the helical (torsional) sector and by the magnetic/flux background.
	Overall, the figure provides a compact characterization of how the Cornell Coulomb parameter controls the depth of the bound states in Model~II.
	
	\begin{figure*}[tbhp]
		\centering
		\includegraphics[width=0.98\linewidth]{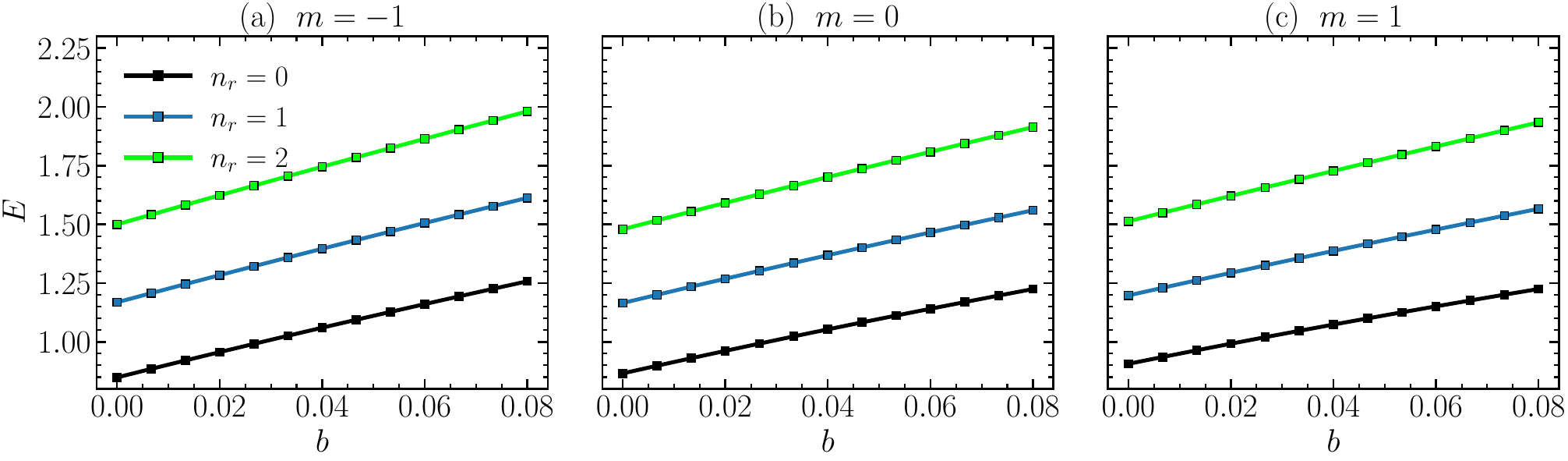}
		\caption{%
			Bound-state energies $E$ as a function of the Cornell linear parameter $b$ in
			$V_{\mathrm C}(r)=a/r+b\,r$ [Eq.~\eqref{eq:Cornell}] for the helically twisted geometry (Model~II).
			Panels (a)--(c) correspond to the fixed azimuthal sectors $m=-1$, $0$, and $1$, respectively.
			In each panel, the three lowest radial levels $n_r=0,1,2$ are displayed as functions of $b$,
			as obtained from the reduced radial equation~\eqref{eq:SL} with Dirichlet boundary conditions~\eqref{eq:BC}.
			The gauge sector is kept on ($B_0=0.5$ and $\Phi_B=0.5$).
			The remaining parameters are fixed to $\hbar=\mu=e=k=1$, $\omega=1$, and $a=1$, while $b$ is varied in $0\le b\le 0.08$
			(dimensionless units).%
		}
		\label{fig:E_vs_b_Cornell}
	\end{figure*}
	
	Figure~\ref{fig:E_vs_b_Cornell} shows how the low-lying spectrum changes as the linear Cornell coupling $b$ is increased at fixed $(\omega,a)$.
	In all three azimuthal sectors, the eigenenergies increase monotonically with $b$, consistent with the fact that the $+\,b\,r$ term raises the large-$r$ part of the potential and therefore pushes the discrete levels upward.
	For each fixed $m$, the radial ordering is preserved throughout the scanned interval,
	$E_{0m}<E_{1m}<E_{2m}$, and the level spacings grow mildly with $b$, reflecting the enhancement of radial quantization produced by the stronger confining tail.
	Comparing panels (a)--(c), the vertical offsets and slopes depend on $m$ because the azimuthal sector couples to the gauge--torsion structure encoded in $V_1(r)$, modifying the balance between the centrifugal-like contribution and the $\omega$-dependent terms induced by the helical distortion and by the magnetic/flux coupling.
	Overall, the figure provides a compact characterization of the sensitivity of the lowest bound states to the Cornell linear coupling in Model~II.
	
	\begin{figure*}[tbhp]
		\centering
		\includegraphics[width=0.98\linewidth]{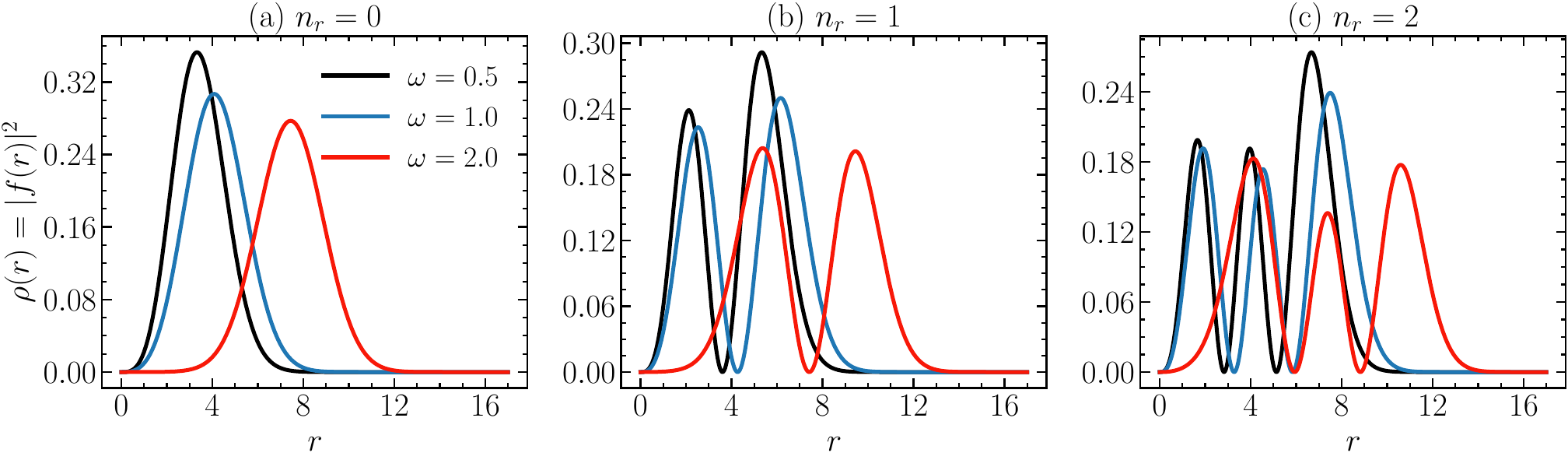}
		\caption{%
			Probability densities $\rho(r)=|f(r)|^{2}$ for the Cornell interaction
			$V_{\mathrm{C}}(r)=a/r+b\,r$ [Eq.~\eqref{eq:Cornell}] in the helically twisted geometry (Model~II).
			Panels (a)--(c) display the three lowest radial modes $n_r=0,1,2$, respectively, in the fixed azimuthal sector $m=1$.
			In each panel, the three curves correspond to $\omega=0.5$, $1.0$, and $2.0$ (as indicated in the legend),
			with $a=1.0$ and $b=0.02$ kept fixed.
			The gauge sector entering the universal contribution $V_1(r)$ is kept on with $B_0=0.5$ and $\Phi_B=0.5$,
			and the remaining parameters are $\hbar=\mu=e=k=1$ (dimensionless units).
			The eigenfunctions are normalized on the numerical domain as
			$\int_{r_{\min}}^{r_{\max}} |f(r)|^{2}\,dr=1$.%
		}
		\label{fig:rho_vs_r_Cornell_omega}
	\end{figure*}
	
	Figure~\ref{fig:rho_vs_r_Cornell_omega} illustrates how the torsion parameter $\omega$ modifies the spatial
	localization of the low-lying bound states in the Cornell model within Model~II.
	As $\omega$ increases, $\rho(r)$ becomes more concentrated and its main weight shifts toward smaller radii,
	indicating an enhancement of the effective radial confinement induced by the $\omega$-dependent terms contained in
	the universal contribution $V_1(r)$.
	For the excited states ($n_r=1,2$), the nodal structure is preserved and the outer lobes are progressively compressed,
	showing a reduction of the characteristic radial scale without altering the number of nodes.
	Overall, larger $\omega$ enhances localization for all three modes while maintaining the expected ordering and
	nodal properties of the bound-state wavefunctions.%
	
	\section{Model III: Kratzer external potential}\label{sec5}
	
	In Model~III, we adopt a Kratzer-type external interaction, which is widely used to model molecular binding and other systems where a Coulomb-like attraction is regularized by a short-distance repulsive core. The radial potential is taken as
	\begin{equation}\label{eq:Kratzer}
		V(r)=-2D\left(\frac{A}{r}-\frac{A^2}{2r^2}\right),
	\end{equation}
	with $A>0$ and $D>0$ \cite{26,27,28,29,30,31,32,33}. This form makes explicit its two key ingredients: an attractive Coulomb-like term ($\propto -1/r$) and a repulsive core ($\propto +1/r^2$). The model is particularly informative in the present helically twisted geometry because Eq.~\eqref{eq:Veff-general} already generates both $1/r$ and $1/r^2$ structures; therefore, the Kratzer parameters provide a controlled handle to tune \emph{independently} the long-range attraction and the short-distance effective barrier on top of the intrinsic torsion- and gauge-induced contributions.
	
	From the numerical standpoint, nothing conceptually changes relative to the previous models: we insert $V(r)$ into $V_{\mathrm{eff}}(r)$, construct $U(r)$, discretize Eq.~\eqref{eq:SL} by finite differences, and extract the low-lying spectrum under the boundary conditions~\eqref{eq:BC} across the desired parameter scans. We do not assume any closed-form quantization rule; the bound-state energies are obtained directly from the eigenvalue problem. The additional $1/r^2$ singularity introduced by the Kratzer core is handled in practice by the inner cutoff $r_{\min}$, together with systematic convergence checks.
	
	Inserting the Kratzer interaction~\eqref{eq:Kratzer} into Eq.~\eqref{eq:Veff-general}, we obtain
	\begin{align}\label{eq:Veff-III}
		V_{\mathrm{eff}}^{(\mathrm{III})}(r)
		&=-\frac{2DA}{r}+\frac{DA^{2}}{r^{2}}
		+\frac{\hbar^{2}}{2\mu}\,%
		\left[
		\frac{m^{2}-\tfrac{1}{4}}{r^{2}}
		+k^{2}(1+\omega^{2})
		-\frac{2m\omega k}{r}
		+\frac{2\omega k e}{r}\,A_{\varphi}(r)
		-\frac{2me}{r^{2}}\,A_{\varphi}(r)
		+\frac{e^{2}}{r^{2}}\,A_{\varphi}^{2}(r)
		\right]\,.
	\end{align}
	Equation~\eqref{eq:Veff-III} also clarifies how the Kratzer parameters reorganize the near-origin structure. The repulsive core $DA^2/r^2$ effectively renormalizes the centrifugal coefficient, so that the net $1/r^2$ sector may be written as
	\[
	\frac{\hbar^2}{2\mu}\left[\frac{m^2-\tfrac14+\left(\tfrac{2\mu}{\hbar^2}\right)DA^2}{r^2}\right],
	\]
	while the attractive Kratzer term $-2DA/r$ adds directly to the torsion-induced Coulomb-like contribution $\propto -1/r$. Consequently, depending on the sign of $m\omega k$, the $1/r$ sector can be enhanced or partially compensated, allowing one to systematically deepen (or soften) the potential well while simultaneously controlling the short-distance barrier through $DA^2$.
	\begin{figure}[!ht]
		\centering
		\includegraphics[width=0.48\linewidth]{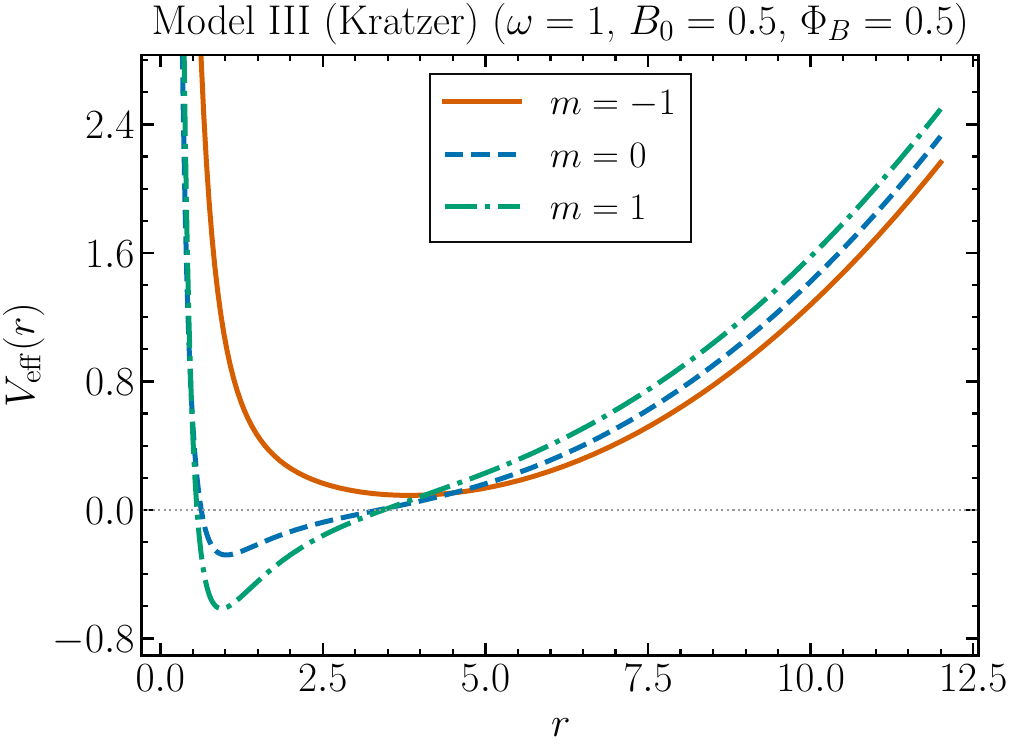}
		\caption{Effective potential $V_{\mathrm{eff}}^{(\mathrm{III})}(r)$
			[Eq.~\eqref{eq:Veff-III}] for Model~III (Kratzer interaction) in the helically twisted geometry
			in the presence of a uniform magnetic field and an Aharonov--Bohm flux.
			The azimuthal vector potential is kept throughout,
			$A_{\varphi}(r)=-(B_0/2)\,r^{2}+\Phi_{B}/(2\pi)$.
			The Kratzer core generates an additional $1/r^{2}$ barrier (through the $A^{2}/r^{2}$ term in $V(r)$),
			which enhances the short--distance repulsion, while the attractive Coulomb--like tail $-2DA/r$ competes with
			the torsion--induced contribution $\propto -2m\omega k/r$ (attractive for $m>0$ and repulsive for $m<0$),
			leading to a pronounced potential well for $m=0$ and, in particular, for $m=1$.
			Parameters: $A=1$, $D=1$, $\hbar=\mu=e=k=1$, $\omega=1$, $B_0=0.5$, and $\Phi_B=0.5$
			(dimensionless units).}
		\label{fig:Veff-III}
	\end{figure}
	
	Figure~\ref{fig:Veff-III} shows $V_{\mathrm{eff}}^{(\mathrm{III})}(r)$ for $m=-1,0,1$.
	Relative to Model~I, the Kratzer interaction introduces a stronger inner barrier and a stronger intermediate--range attraction.
	At short distances, the effective $1/r^{2}$ wall is steepened by the combined centrifugal term and the Kratzer core
	(which contributes an extra $A^{2}/r^{2}$ component through $V(r)$), pushing the inner turning point outward.
	At intermediate radii, the attractive Kratzer tail $-2DA/r$ deepens the well; for $m=1$ this attraction cooperates with the
	torsion--induced Coulomb--like term $\propto -2m\omega k/r$, yielding the deepest minimum among the three sectors,
	whereas for $m=-1$ the torsion term is repulsive and partially compensates the Kratzer attraction.
	At large $r$, the rise of $V_{\mathrm{eff}}^{(\mathrm{III})}(r)$ is dominated by the magnetic confinement encoded in $A_\varphi$,
	producing the characteristic parabolic outer wall.
	Overall, the Kratzer coupling enhances radial confinement and increases the binding tendency compared to Model~I,
	while preserving the $m$--dependent splitting controlled by the geometric and electromagnetic contributions.

	\begin{figure*}[tbhp]
		\centering
		\includegraphics[width=0.98\linewidth]{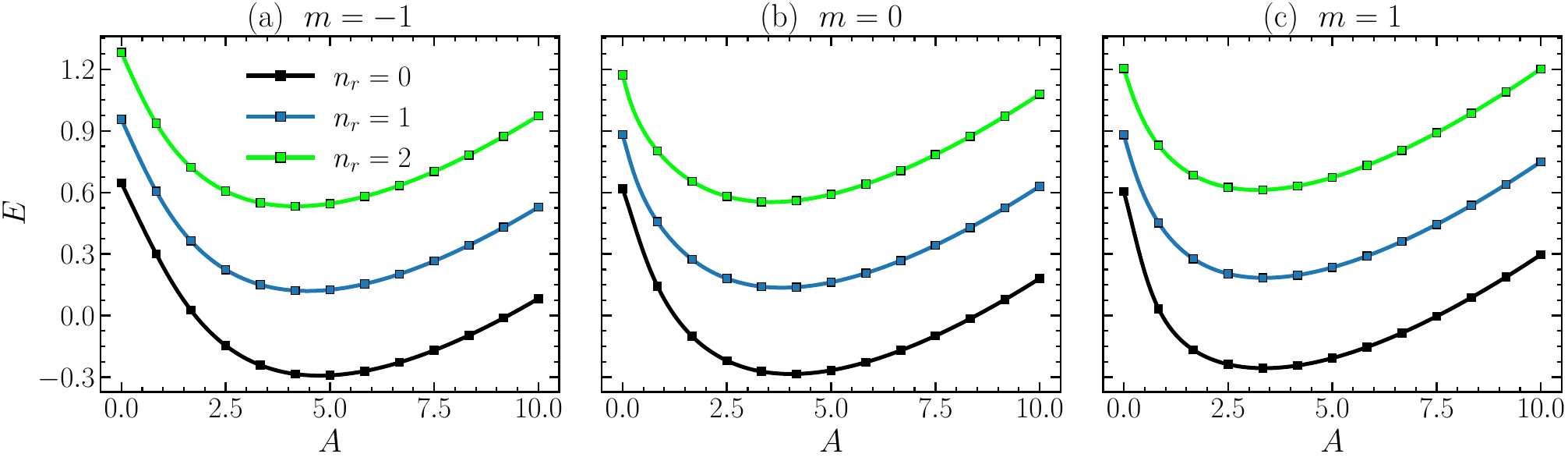}
		\caption{%
			Bound-state energies $E$ as a function of the Kratzer parameter $A$ for the external interaction $V_{\mathrm K}(r)$ defined in Eq.~\eqref{eq:Kratzer} (Model~III).
			The panels correspond to fixed azimuthal sectors: (a) $m=-1$, (b) $m=0$, and (c) $m=1$.
			In each panel we display the three lowest radial levels $n_r=0,1,2$ obtained from the reduced Sturm--Liouville equation~\eqref{eq:SL} with Dirichlet boundary conditions~\eqref{eq:BC}.
			The parameters are fixed to $\hbar=\mu=e=k=1$, $\omega=1$, $D=1$, with the gauge sector kept on ($B_0=0.5$ and $\Phi_B=0.5$) (dimensionless units).%
		}
		\label{fig:E_vs_A_ModelIII_Kratzer}
	\end{figure*}
	Figure~\ref{fig:E_vs_A_ModelIII_Kratzer} shows that the low-lying spectrum exhibits a non-monotonic dependence on the Kratzer parameter $A$.
	For small $A$, the energies decrease as the attractive Coulomb-like contribution $\propto -A/r$ deepens the effective radial well.
	However, increasing $A$ simultaneously enhances the repulsive short-distance term $\propto +A^{2}/r^{2}$, which stiffens the inner region of the potential; as a result, each branch reaches a minimum and then rises for larger $A$.
	The positions of the minima and the overall vertical offsets differ among panels (a)--(c), reflecting the $m$-dependent coupling to the torsion--gauge structure encoded in $V_1(r)$ [Eq.~\eqref{eq:V1}], which modifies both the Coulomb-like and centrifugal-like contributions in the effective problem.
	Across the scanned interval, the ordering of the radial modes is preserved, with $n_r=0$ as the ground state followed by $n_r=1$ and $n_r=2$. 
	
	\begin{figure*}[tbhp]
		\centering
		\includegraphics[width=\linewidth]{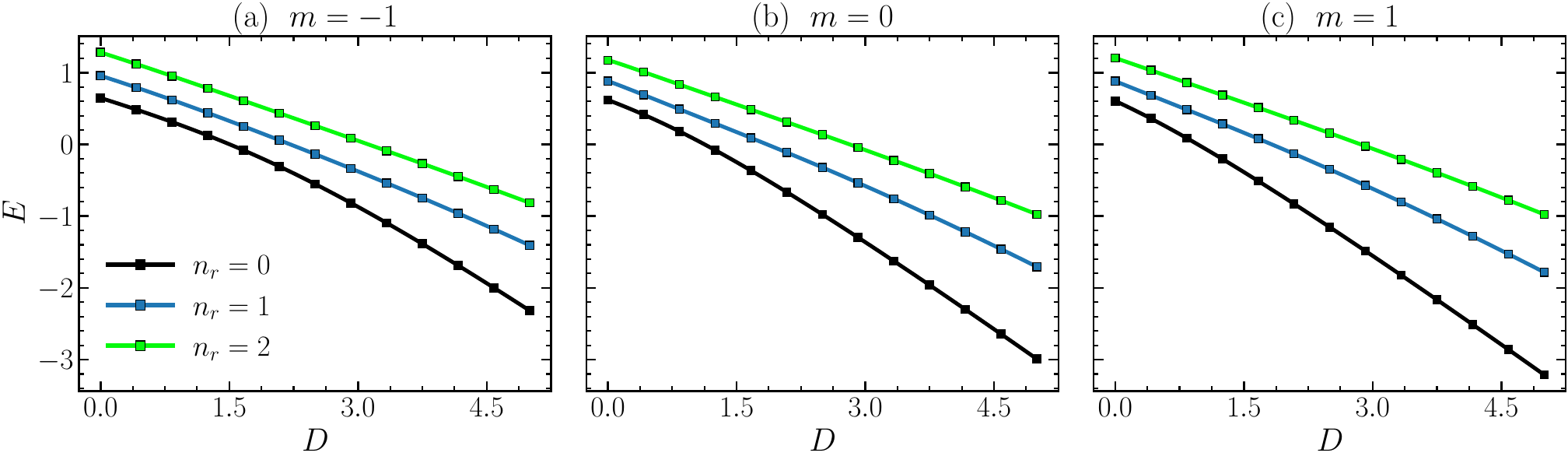}
		\caption{
			Bound-state energies $E$ as a function of the Kratzer strength $D$ for the interaction in Eq.~\eqref{eq:Kratzer}, with $A=1$.
			Panels (a)--(c) correspond to the azimuthal sectors (a) $m=-1$, (b) $m=0$, and (c) $m=1$.
			In each panel, the three lowest radial levels $n_r=0,1,2$ are displayed.
			The remaining parameters are fixed to $\hbar=\mu=1$, $k=1$, and $\omega=1$ (dimensionless units).
		}
		\label{fig:E_vs_D_model3}
	\end{figure*}
	Figure~\ref{fig:E_vs_D_model3} displays the dependence of the low-lying spectrum on the Kratzer coupling $D$ for fixed $A$ and $\omega$.
	As $D$ increases, all branches shift downward, which is consistent with a net strengthening of the attractive part of the Kratzer interaction and, therefore, with more strongly bound states.
	For each fixed $m$, the level ordering is preserved throughout the scanned interval,
	$E_{0m}<E_{1m}<E_{2m}$,
	with the ground state ($n_r=0$) being the most sensitive to variations of $D$.
	Comparing panels (a)--(c), the curves exhibit an overall $m$-dependent offset, reflecting how the Kratzer terms combine with the $m$-dependent contributions already present in the reduced radial equation.
	Overall, the figure provides a clear characterization of how the Kratzer strength controls both the depth of the bound states and the spacing among the lowest radial levels in Model~III.
	
	\begin{figure*}[tbhp]
		\centering
		\includegraphics[width=\linewidth]{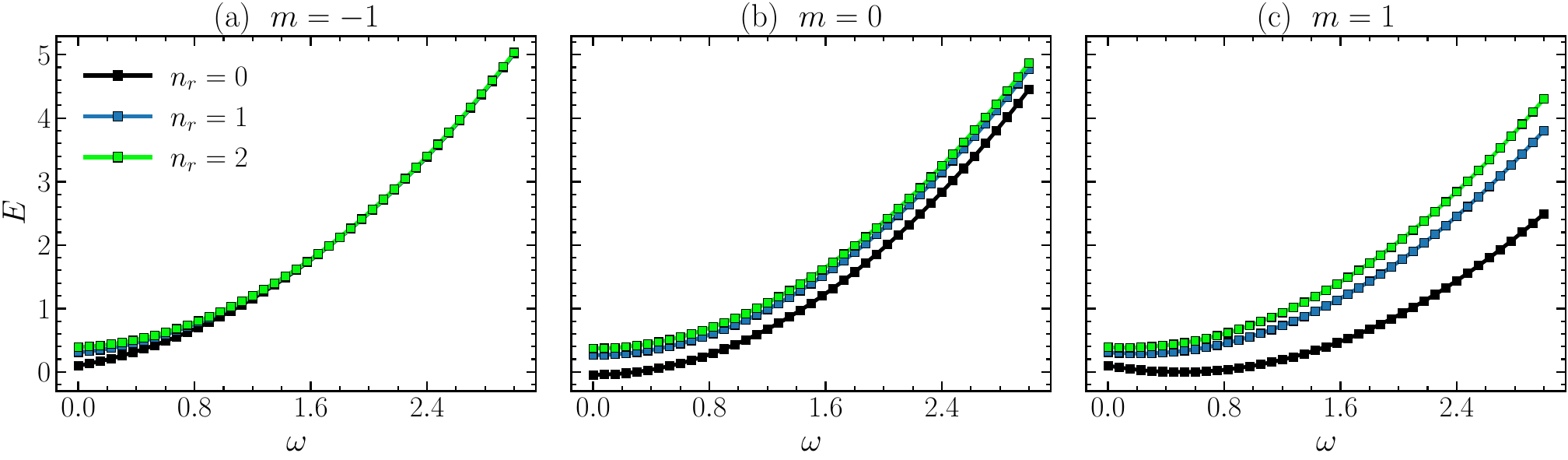}
		\caption{%
			Bound-state energies $E$ as a function of the torsion parameter $\omega$ for the Kratzer interaction $V_{\mathrm{K}}(r)$ given in Eq.~\eqref{eq:Kratzer} (Model~III).
			The panels correspond to fixed azimuthal sectors:
			(a) $m=-1$, (b) $m=0$, and (c) $m=1$.
			In each panel we display the three lowest radial levels $n_r=0,1,2$ obtained from the reduced Sturm--Liouville problem~\eqref{eq:SL}
			with Dirichlet boundary conditions~\eqref{eq:BC}.
			The remaining parameters are fixed to $\hbar=\mu=1$, $k=1$, and $A=D=1$ (dimensionless units).%
		}
		\label{fig:E_vs_omega_model3}
	\end{figure*}
	Figure~\ref{fig:E_vs_omega_model3} displays the dependence of the low-lying spectrum on the torsion parameter $\omega$ in the Kratzer model with $A=D=1$.
	For all azimuthal sectors $m=-1,0,1$, the energies increase monotonically as $\omega$ grows, showing that the $\omega$-dependent geometric contribution shifts the spectrum upward in the parameter range considered.
	For each fixed $(m,\omega)$, the radial ordering is preserved, with the ground state $n_r=0$ lying below the excited levels $n_r=1$ and $n_r=2$.
	Moreover, the spacing between adjacent radial levels tends to increase with $\omega$, indicating an enhanced sensitivity of the bound states to torsion when the Kratzer interaction is present, with a noticeable $m$-dependent splitting across panels (a)--(c).
	
	\begin{figure*}[tbhp]
		\centering
		\includegraphics[width=0.98\linewidth]{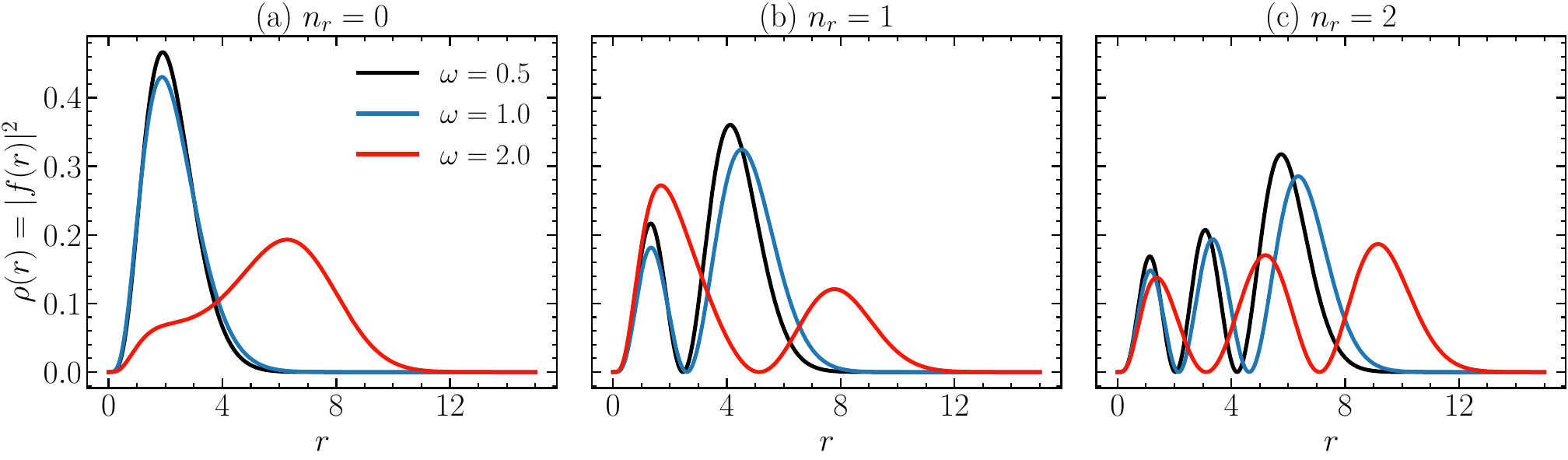}
		\caption{%
			Probability densities for the Kratzer model in the helically twisted geometry.
			Panels (a)--(c) show the lowest radial modes $n_r=0,1,2$, respectively.
			In each panel, the curves correspond to $\omega=0.5$, $1.0$, and $2.0$, while all other parameters are kept fixed as in the corresponding spectral scan (including the Kratzer couplings, the azimuthal sector, and the gauge parameters).
			The eigenfunctions are normalized as $\int_{r_{\min}}^{r_{\max}} |f(r)|^{2}\,dr=1$ (dimensionless units).%
		}
		\label{fig:rho_vs_r_ModelIII}
	\end{figure*}
	Figure~\ref{fig:rho_vs_r_ModelIII} illustrates how the helical--torsion parameter $\omega$ reshapes the spatial profiles of the low-lying bound states in the Kratzer model.
	For the ground state $n_r=0$ [Fig.~\ref{fig:rho_vs_r_ModelIII}(a)], varying $\omega$ changes both the peak position and the degree of localization.
	In particular, when the gauge sector is kept on, the $\omega$--dependent coupling to $A_\varphi(r)$ introduces an additional radial structure in the effective interaction that can shift the dominant probability weight to a finite-radius region, rather than monotonically concentrating it toward the origin.
	For the excited states $n_r=1$ and $n_r=2$ [Figs.~\ref{fig:rho_vs_r_ModelIII}(b) and \ref{fig:rho_vs_r_ModelIII}(c)], the nodal pattern is preserved, while the locations and relative heights of the lobes are noticeably modified as $\omega$ varies, reflecting the competition between the Kratzer terms and the $\omega$-dependent gauge--geometric contributions.
	Overall, the profiles confirm that $\omega$ provides an efficient control parameter for the radial localization scale and for the redistribution of probability across the low-lying modes.%

	\section{Model IV: Small-oscillation regime of the Morse interaction}\label{sec6}
	
	In the fourth scenario we incorporate an external interaction that captures
	\emph{vibrational} confinement near an equilibrium radius. For this purpose,
	we adopt the Morse potential, which is a standard model for small oscillations
	around a stable minimum and, when expanded about equilibrium, naturally yields
	an effective quadratic-plus-linear form. In the present helically twisted
	geometry, this approximation is useful because it allows a transparent
	separation between: (i) confinement generated intrinsically by the torsion/gauge
	background through $V_{\mathrm{eff}}(r)$, and (ii) an \emph{explicit} nearly
	harmonic confinement introduced externally.
	
	We start from the Morse interaction \cite{34,35,36}
	\begin{equation}\label{eq:Morse}
		V(r)= D\left[1-e^{-a(r-r_0)}\right]^2 - D,
	\end{equation}
	where $D>0$ is the dissociation energy scale, $a>0$ controls the stiffness, and
	$r_0$ is the equilibrium position. The potential has a stable minimum at
	$r=r_0$, consistent with the small-oscillation conditions \cite{13}
	\begin{equation}\label{eq:Morse_conditions}
		\left.\frac{dV}{dr}\right|_{r=r_0}=0,\qquad
		\left.\frac{d^2V}{dr^2}\right|_{r=r_0}>0.
	\end{equation}
	Expanding Eq.~\eqref{eq:Morse} up to second order about $r=r_0$ leads to the
	quadratic-plus-linear approximation
	\begin{equation}\label{eq:Morse_small}
		V(r)\approx D a^2 r^2-2(Da^2 r_0)\,r + D_1,\qquad D_1=D(a^2 r_0^2-1).
	\end{equation}
	In this model we take Eq.~\eqref{eq:Morse_small} as the external interaction.
	This choice isolates the near-equilibrium vibrational physics while keeping the
	full geometric and magnetic/flux-induced structures already present in
	$V_{\mathrm{eff}}(r)$.
	
	As in the previous models, the numerical pipeline is unchanged: once $V(r)$ is
	fixed, we build $V_{\mathrm{eff}}(r)$ and $U(r)$ via
	Eqs.~\eqref{eq:Veff-general} and \eqref{eq:U_def}, and then solve the
	Sturm--Liouville problem~\eqref{eq:SL} under the boundary conditions
	\eqref{eq:BC}. Parameter scans are performed by repeating the same
	finite-difference discretization and tridiagonal diagonalization.
	
	With the small-oscillation approximation~\eqref{eq:Morse_small}, the effective
	potential becomes
	\begin{align}\label{eq:Veff-IV}
		V_{\mathrm{eff}}^{(\mathrm{IV})}(r)
		&=Da^{2}r^{2}-2Da^{2}r_{0}\,r+D(a^{2}r_{0}^{2}-1)\notag\\&
		+\frac{\hbar^{2}}{2\mu}\,%
		\left[
		\frac{m^{2}-\tfrac{1}{4}}{r^{2}}
		+k^{2}(1+\omega^{2})
		-\frac{2m\omega k}{r}
		+\frac{2\omega k e}{r}\,A_{\varphi}(r)
		-\frac{2me}{r^{2}}\,A_{\varphi}(r)
		+\frac{e^{2}}{r^{2}}\,A_{\varphi}^{2}(r)
		\right]\,.
	\end{align}
	Equation~\eqref{eq:Veff-IV} makes the role of each sector explicit. The
	quadratic external term $Da^2\,r^2$ reinforces the magnetic parabolic
	confinement $\propto r^2$, so that the overall large-$r$ curvature is enhanced
	relative to Model~I. Meanwhile, the linear pieces $-2Da^2 r_0\,r$ and
	$-(\hbar^2/2\mu)\,\omega k eB_0\,r$ primarily shift the equilibrium position
	without destroying the confining character of the potential, thereby enabling a
	controlled displacement of the minimum.
	
	\begin{figure}[!ht]
		\centering
		\includegraphics[width=0.48\linewidth]{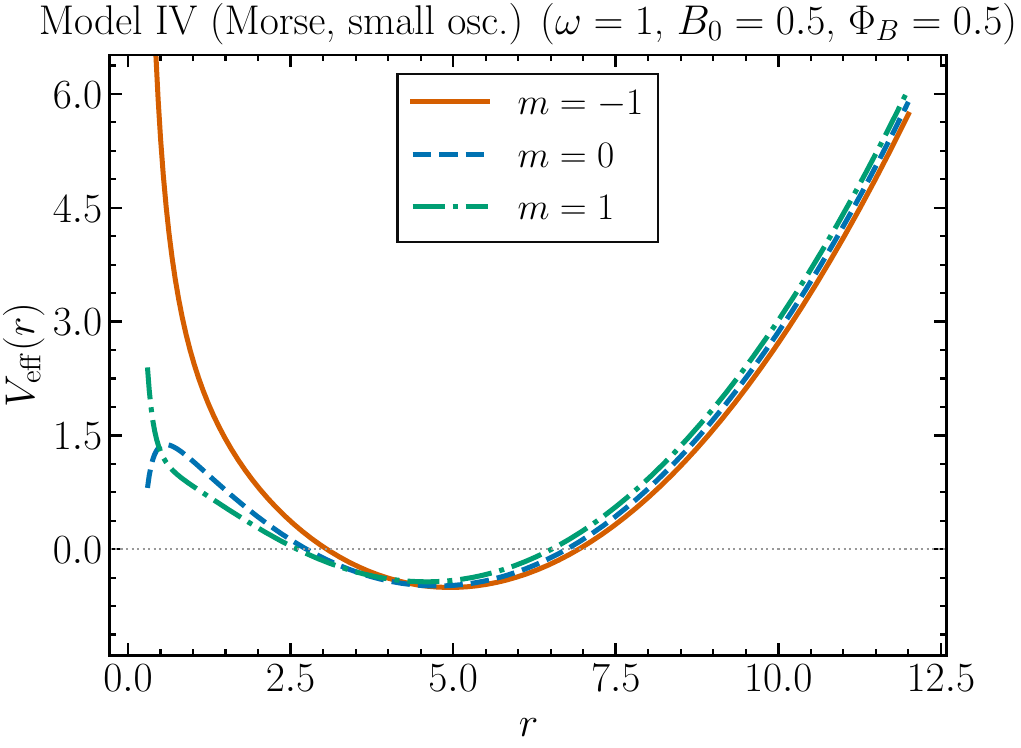}
		\caption{Effective potential $V_{\mathrm{eff}}^{(\mathrm{IV})}(r)$
			[Eq.~\eqref{eq:Veff-IV}] for Model~IV (Morse interaction in the small--oscillation regime)
			in the helically twisted geometry in the presence of a uniform magnetic field and an Aharonov--Bohm flux.
			The Morse potential is expanded about its minimum $r_0$, yielding
			$V(r)\simeq D a^{2}r^{2}-2Da^{2}r_0\,r + D(a^{2}r_0^{2}-1)$.
			The azimuthal vector potential is kept throughout,
			$A_{\varphi}(r)=-(B_0/2)\,r^{2}+\Phi_{B}/(2\pi)$, so that the magnetic confinement and the AB shift are fully included.
			The external quadratic term $Da^{2}r^{2}$ reinforces the magnetic outer wall, while the linear term $-2Da^{2}r_0\,r$
			shifts the position of the minimum and competes with the $m$--dependent Coulomb--like geometric contribution.
			Parameters: $D=1$, $a=0.30$, $r_0=5$, $\hbar=\mu=e=k=1$, $\omega=1$, $B_0=0.5$, and $\Phi_B=0.5$
			(dimensionless units).}
		\label{fig:Veff-IV}
	\end{figure}
	
	Figure~\ref{fig:Veff-IV} displays $V_{\mathrm{eff}}^{(\mathrm{IV})}(r)$ for $m=-1,0,1$.
	Among the four models, Model~IV produces the strongest confinement because two independent mechanisms build the outer wall:
	(i) the magnetic contribution encoded in $A_{\varphi}$ and (ii) the quadratic term $Da^{2}r^{2}$ from the small--oscillation expansion.
	As a result, the effective potential develops a pronounced well in all azimuthal sectors for the parameters shown.
	The location of the minimum is controlled primarily by the linear Morse term $-2Da^{2}r_0\,r$, while the relative depth and
	left--right shift of the well depend on $m$ through the geometric $1/r$ coupling and the $A_\varphi$ terms in Eq.~\eqref{eq:Veff-IV}.
	In particular, the $m=1$ sector (where the geometric Coulomb--like term is attractive) exhibits the most compact and deepest well,
	whereas for $m=-1$ the geometric contribution is repulsive and the minimum is shifted to larger radii and becomes shallower.
	Overall, Model~IV provides a highly tunable confining landscape: the well depth, width, and equilibrium radius can be adjusted by
	$(D,a,r_0)$ in addition to the geometric parameters $(\omega,k)$ and the electromagnetic set $(B_0,\Phi_B)$.
	
	\begin{figure}[tbhp]
		\centering
		\includegraphics[width=0.45\linewidth]{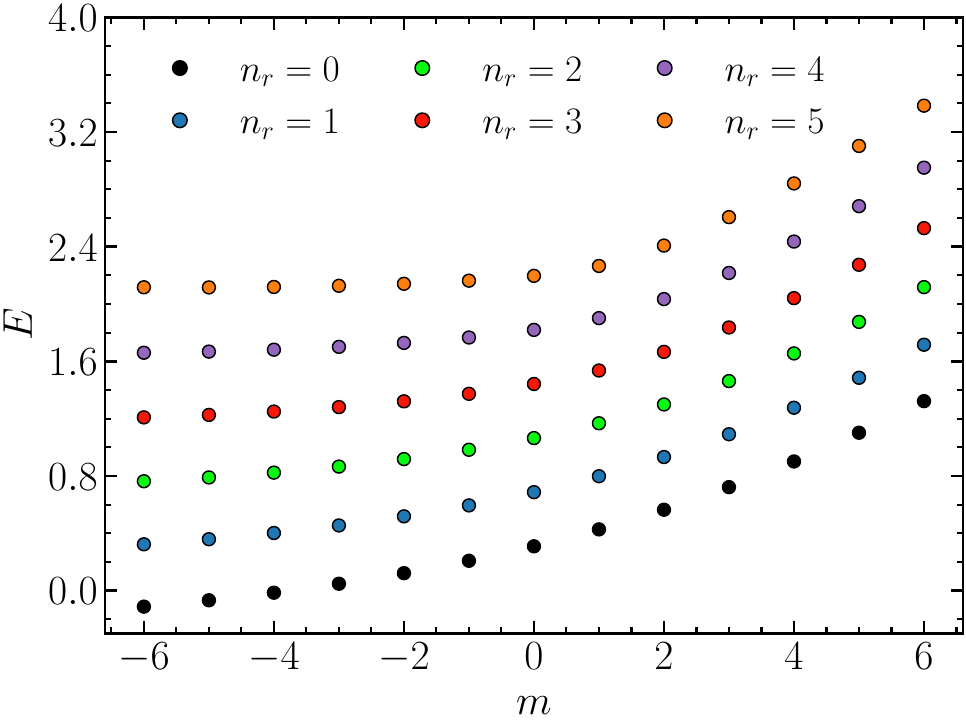}
		\caption{
			Numerical energy levels $E_{n_r m}$ as a function of the azimuthal quantum number $m$ for the small-oscillation (Morse) model in the helically twisted geometry.
			The external interaction is taken as the quadratic-plus-linear approximation obtained from the expansion of the Morse potential around its minimum [Eq.~\eqref{eq:Morse_small}].
			The physical energies are obtained from the eigenvalues as $E=(\hbar^{2}/2\mu)\lambda$.
			The effective potential includes the vector-potential sector through $A_{\varphi}(r)$ (i.e., $B_0$ and the Aharonov--Bohm flux $\Phi_B$ are kept on), consistently with the general definition of $V_{\mathrm{eff}}(r)$ in Eq.~\eqref{eq:Veff-general}.
			We display the six lowest radial levels $n_r=0,1,2,3,4,5$.
			Parameters are fixed to $\hbar=\mu=k=1$ (dimensionless units), $\omega=1$, and $(D,a,r_0)=(1,0.2,10)$ in Eq.~\eqref{eq:Morse_small}.
		}
		\label{fig:E_vs_m_Morse}
	\end{figure}
	Figure~\ref{fig:E_vs_m_Morse} shows the discrete spectrum obtained by direct diagonalization of the finite-difference representation of the reduced Sturm--Liouville operator.
	For each fixed $m$, the states form a ladder labeled by the radial index $n_r$, with $E_{0m}<E_{1m}<\cdots<E_{5m}$, as expected for a confining small-oscillation potential.
	The dependence on $m$ is not symmetric under $m\to -m$, reflecting the presence of terms in $U(r)$ that are linear in $m$ (originating from the helical-torsion coupling and from the vector-potential sector through $A_{\varphi}$), in addition to the centrifugal-like contribution $(m^{2}-1/4)/r^{2}$.
	As $m$ increases, all branches shift upward and their spacing changes smoothly, indicating that the azimuthal sector controls both the effective short-distance barrier and the strength of the $m$-dependent geometric/electromagnetic couplings.
	
	\begin{figure*}[tbhp]
		\centering
		\includegraphics[width=0.98\linewidth]{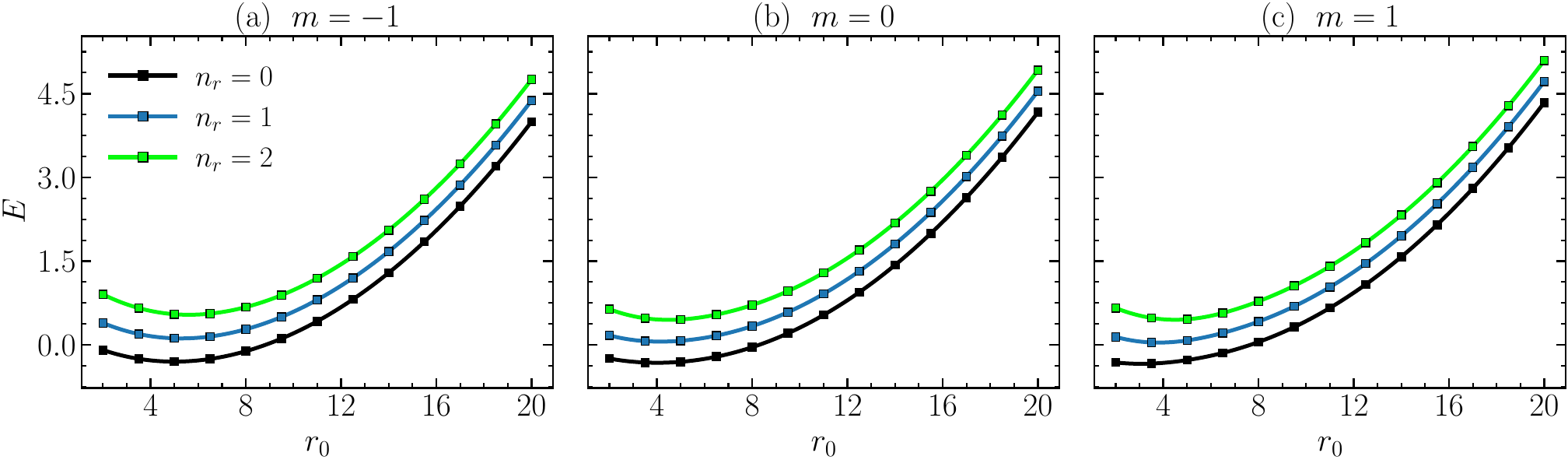}
		\caption{%
			Numerical bound-state energies $E$ as a function of the equilibrium position parameter $r_{0}$ entering the small-oscillation (quadratic-plus-linear) approximation of the Morse interaction [Eq.~\eqref{eq:Morse_small}], for the three lowest radial modes $n_r=0,1,2$.
			The three panels correspond to fixed azimuthal sectors: (a) $m=-1$, (b) $m=0$, and (c) $m=1$.
			The reduced radial Sturm--Liouville problem is solved on a finite interval with Dirichlet boundary conditions and discretized by a second-order finite-difference scheme, yielding a real symmetric tridiagonal matrix.
			Eigenvalues $\lambda$ are converted to energies through $E=(\hbar^{2}/2\mu)\lambda$ with $\hbar=\mu=k=1$.
			The geometric parameter is fixed to $\omega=1$, and the vector-potential coupling is kept throughout via
			$A_{\varphi}(r)=-(B_0/2)\,r^{2}+\Phi_B/(2\pi)$ with $B_0=0.5$ and $\Phi_B=0.5$ (and $e=1$).
			The Morse parameters are fixed at $D=1$ and $a=0.2$, while $r_{0}$ is varied.%
		}
		\label{fig:E_vs_r0_Morse}
	\end{figure*}
	Figure~\ref{fig:E_vs_r0_Morse} shows how shifting the equilibrium position $r_{0}$ in the small-oscillation approximation of the Morse interaction reshapes the low-lying spectrum when the vector-potential coupling is kept on.
	Varying $r_{0}$ changes the coefficient of the linear term in the effective confinement generated by the Morse expansion, which effectively displaces the minimum of the radial potential in the reduced Sturm--Liouville operator.
	As a consequence, the energy branches display a smooth $r_{0}$-dependence whose detailed profile results from the competition between: (i) the $r_{0}$-dependent linear contribution in $V_{\rm ext}(r)$, (ii) the geometric torsion terms controlled by $\omega$, and (iii) the magnetic/flux contributions entering through $A_\varphi(r)$.
	
	For the three azimuthal sectors shown, the radial ordering is preserved at fixed $(m,r_0)$, with
	$E_{n_r=0}<E_{n_r=1}<E_{n_r=2}$.
	Comparing panels (a)--(c), changing $m$ mainly produces an overall vertical displacement and slight curvature changes due to the $m$-dependent terms in $U(r)$, including the centrifugal-like pieces and the couplings involving $A_\varphi$.
	In the large-$r_{0}$ regime, the curves tend to approach a slowly varying behavior, indicating that increasing $r_{0}$ predominantly shifts the effective potential minimum rather than substantially altering the local curvature that controls the level spacing.

	\begin{figure*}[tbhp]
		\centering
		\includegraphics[width=0.98\linewidth]{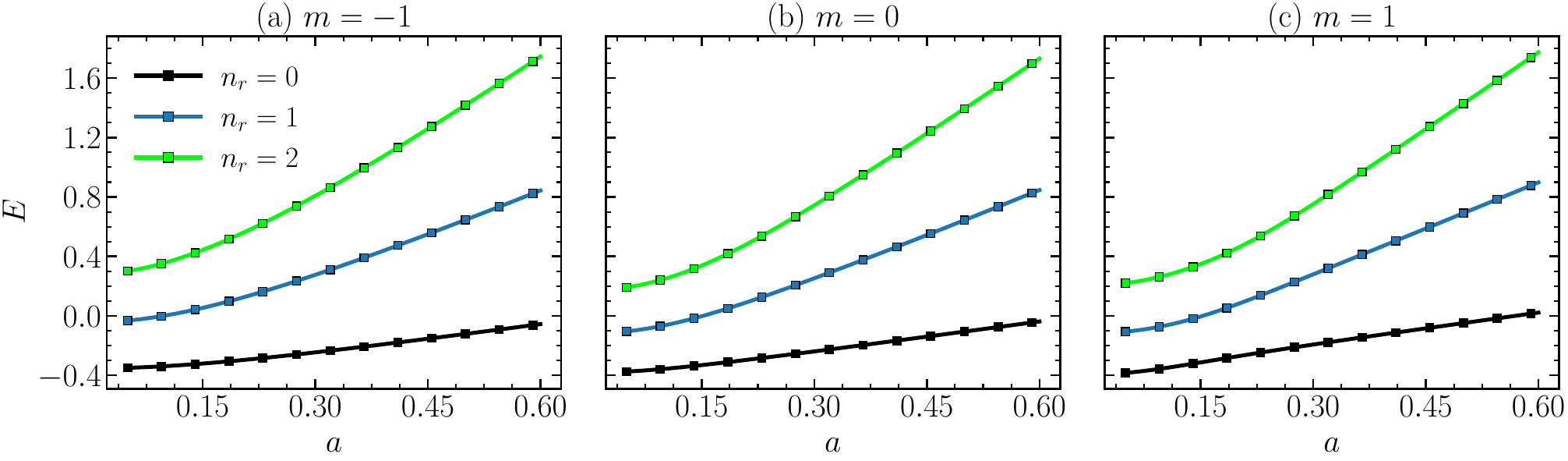}
		\caption{%
			Numerical bound-state energies $E$ as a function of the Morse parameter $a$ in the small-oscillation approximation of the Morse interaction [Eq.~\eqref{eq:Morse_small}], for the three lowest radial modes $n_r=0,1,2$.
			The three panels correspond to fixed azimuthal sectors: (a) $m=-1$, (b) $m=0$, and (c) $m=1$.
			The reduced radial Sturm--Liouville problem is discretized on a finite interval using a second-order finite-difference scheme with Dirichlet boundary conditions, yielding a real symmetric tridiagonal matrix.
			Eigenvalues $\lambda$ are converted to energies through $E=(\hbar^{2}/2\mu)\lambda$ with $\hbar=\mu=k=1$.
			The geometric parameter is fixed to $\omega=1$ and the vector-potential coupling is kept throughout via
			$A_{\varphi}(r)=-(B_0/2)\,r^{2}+\Phi_B/(2\pi)$ with $B_0=0.5$ and $\Phi_B=0.5$ (and $e=1$).
			All remaining Morse parameters are fixed (in particular $D=1$ and $r_{0}=5$), so that the curves isolate the effect of $a$ on the effective curvature (quadratic confinement) generated by the Morse expansion.%
		}
		\label{fig:E_a_Morse}
	\end{figure*}
	Figure~\ref{fig:E_a_Morse} shows the dependence of the low-lying spectrum on the Morse parameter $a$ within the small-oscillation approximation, with the vector-potential coupling kept on.
	Since $a$ controls the local curvature of the Morse well around its equilibrium position, increasing $a$ produces a stiffer effective confinement (the quadratic contribution scales as $\sim D a^{2}r^{2}$), and consequently all bound-state energies shift upward.
	This monotonic increase with $a$ is observed in all azimuthal sectors and becomes more pronounced for higher radial excitations: while the ground branch $n_r=0$ rises more moderately, the excited branches $n_r=1$ and $n_r=2$ exhibit a steeper growth as $a$ increases.
	At fixed $(m,a)$ the ordering of the radial levels is preserved as
	$E_{n_r=0}<E_{n_r=1}<E_{n_r=2}$, consistent with the stronger kinetic contribution associated with higher radial excitations in the discretized Sturm--Liouville spectrum.
	Comparing panels (a)--(c), changing $m$ mainly introduces an overall vertical displacement and mild changes in curvature through the $m$-dependent terms in $U(r)$ (including centrifugal-like and torsion-induced contributions, as well as the couplings to $A_\varphi$), while the qualitative $a$-driven stiffening of the confinement remains robust across the three sectors.

	\begin{figure*}[tbhp]
		\centering
		\includegraphics[width=\linewidth]{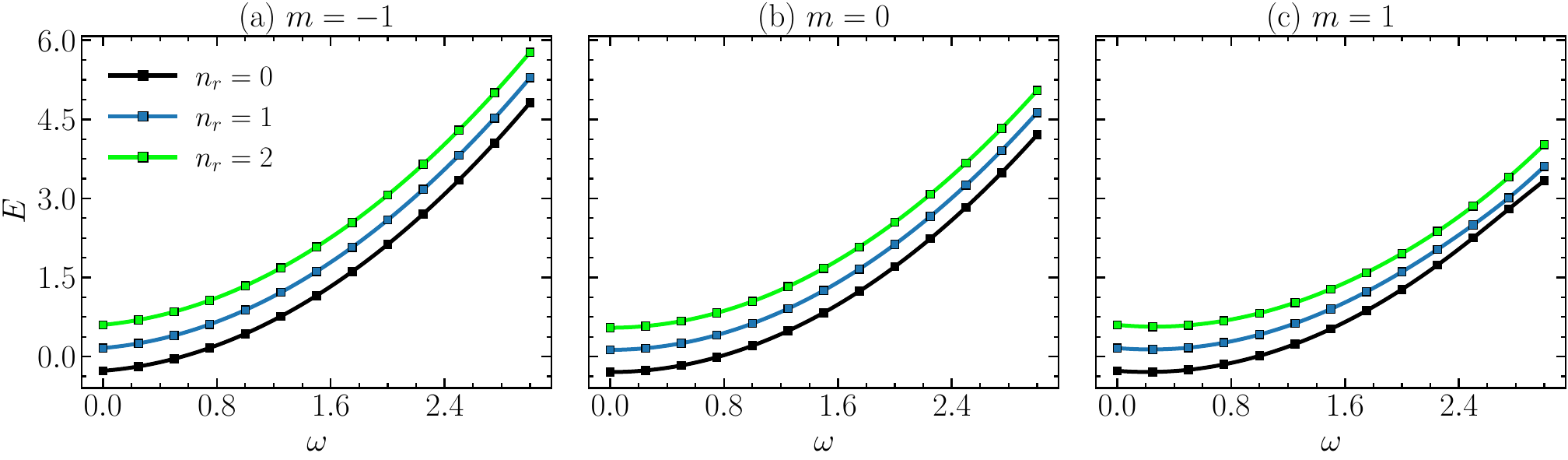}
		\caption{%
			(Morse potential in the small--oscillation regime) Numerical energies $E$ as a function of the geometric torsion parameter $\omega$ for the three lowest radial levels $n_r=0,1,2$ in three representative azimuthal sectors:
			(a) $m=-1$, (b) $m=0$, and (c) $m=1$.
			The Morse interaction is treated in the harmonic (small--oscillation) approximation about $r_0$, with fixed parameters $(D,a,r_0)=(1,0.30,5)$ (dimensionless units).
			Energies are reported with the convention $E=(\hbar^{2}/2\mu)\lambda$, with $\hbar=\mu=k=1$, $B_0=0.5$ and $\Phi_B=0.5$.}
		\label{fig:model4_morse_E_vs_omega}
	\end{figure*}
	Figure~\ref{fig:model4_morse_E_vs_omega} shows the evolution of the low--lying spectrum with the torsion parameter $\omega$ for the Morse interaction in the small--oscillation regime, with the vector potential kept on.
	In all three azimuthal sectors, the eigenenergies increase monotonically with $\omega$.
	This behavior follows from the geometric structure of the effective radial operator: besides the $\omega$--dependent terms coupled to $A_\varphi$, the dominant geometric contribution contains a positive shift $\propto k^2(1+\omega^2)$, which drives the spectrum upward as $\omega$ grows.
	At fixed $(m,\omega)$ the radial ordering is preserved, with the ground state being the lowest branch and the excited levels lying above it, i.e.,
	$E_{n_r=0}<E_{n_r=1}<E_{n_r=2}$.
	Comparing panels (a)--(c), changing $m$ produces an overall displacement and a mild modification of the curvature, while the qualitative monotonic growth with $\omega$ remains robust across the three sectors.
	
	\begin{figure*}[tbhp]
		\centering
		\includegraphics[width=\linewidth]{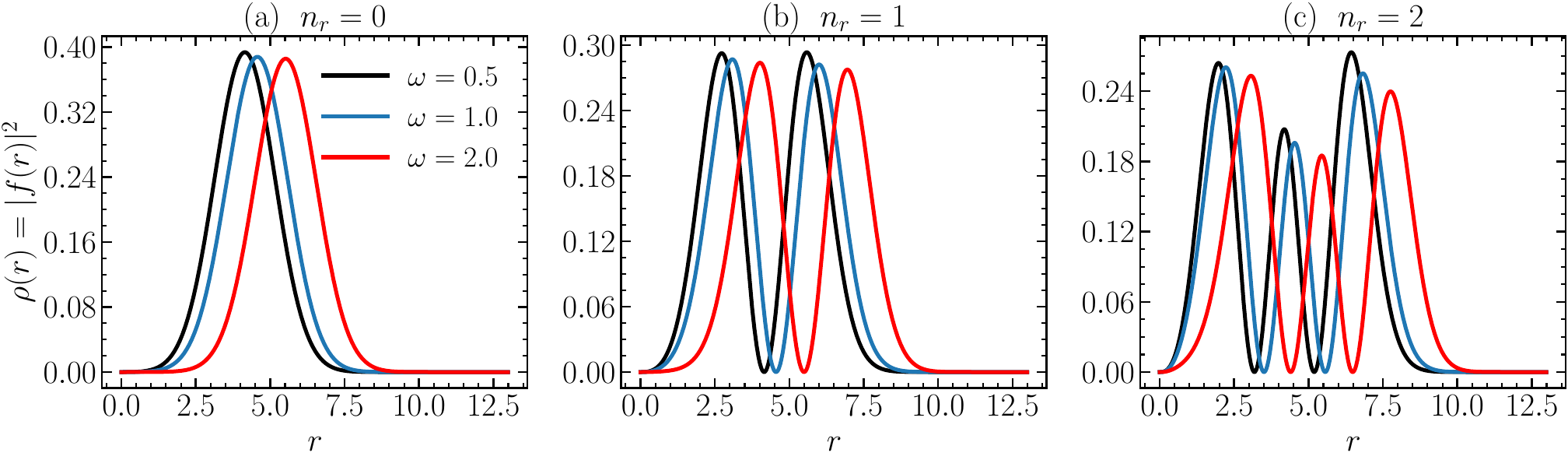}
		\caption{%
			(Model~4: Morse potential in the small--oscillation regime with vector coupling.)
			Radial probability density $\rho(r)=|f(r)|^{2}$ for three torsional frequencies
			$\omega=0.5,\,1.0,$ and $2.0$ (black, blue, and red curves, respectively).
			The panels correspond to the three lowest radial modes:
			(a) $n_r=0$, (b) $n_r=1$, and (c) $n_r=2$, computed at fixed azimuthal sector $m=1$.
			The Morse interaction is treated in the harmonic (small--oscillation) approximation about $r_0$,
			and the vector potential is kept throughout via $A_{\varphi}(r)$ with the fixed parameters
			$B_0=0.5$ and $\Phi_B=0.5$ (dimensionless units).
			The eigenfunctions are obtained from the reduced radial Sturm--Liouville problem using a
			second--order finite--difference discretization with Dirichlet boundary conditions on a truncated domain,
			and are normalized according to $\int_{r_{\min}}^{r_{\max}}|f(r)|^2\,dr=1$.%
		}
		\label{fig:rho_morse_small_omega}
	\end{figure*}
	
	Figure~\ref{fig:rho_morse_small_omega} illustrates how the torsional frequency $\omega$
	reshapes the spatial localization of the bound states in the Morse small--oscillation regime.
	For each fixed radial mode $n_r$, varying $\omega$ shifts the position of the dominant peaks and
	changes the effective width of $\rho(r)$, reflecting the $\omega$--dependence of the geometric sector
	(including the terms coupled to $A_\varphi$).
	The ground state ($n_r=0$) remains single--peaked, while the excited states ($n_r=1,2$) exhibit
	multi--peaked profiles associated with the nodes of the radial wave function.
	We fix $m=1$ to keep the same azimuthal sector across all panels and to highlight the sensitivity of the
	probability density to $\omega$ when the full geometric/vector coupling is retained.%
	
	
	\section{Conclusions}\label{sec7}
	
	We have presented a fully numerical investigation of bound states for a non-relativistic scalar particle in the helically twisted (torsional) three-dimensional geometry described by Eq.~\eqref{eq:metric}, with optional minimal coupling to a uniform magnetic field and an Aharonov--Bohm flux through Eq.~\eqref{eq:Avec}. By separating variables with the ansatz~\eqref{eq:sep}, the problem reduces to a one-dimensional radial Sturm--Liouville eigenvalue equation~\eqref{eq:SL}, whose effective potential $V_{\mathrm{eff}}(r)$, Eq.~\eqref{eq:Veff-general}, naturally contains torsion-induced Coulomb-like and centrifugal-like structures, together with magnetic/flux-dependent contributions and (when present) external interactions. This formulation provides a unified framework in which geometry, gauge fields, and model potentials enter on equal footing through the same self-adjoint operator.
	
	Since analytic special-function constructions and associated truncation prescriptions are not robust across the parameter regimes required for systematic scans, all spectra and eigenfunctions reported here were obtained numerically via a second-order finite-difference discretization of the radial operator on a truncated domain, enforcing Dirichlet boundary conditions~\eqref{eq:BC}. Numerical reliability was ensured by explicit convergence checks under grid refinement, reduction of the inner cutoff, and enlargement of the outer boundary. A practical advantage of this approach is that the discretized operator is real, symmetric, and tridiagonal, allowing efficient and accurate extraction of the low-lying eigenpairs and enabling extensive parameter sweeps without additional approximations beyond controlled discretization errors.
	
	Within this numerical-only strategy, we analyzed four representative scenarios: (i) the purely geometric case $V(r)=0$, (ii) Cornell-type confinement, (iii) Kratzer interaction, and (iv) the small-oscillation (quadratic-plus-linear) approximation of the Morse potential. In all cases, the torsion parameter $\omega$ plays a central role by reshaping the effective radial interaction: increasing $\omega$ systematically shifts the low-lying levels and modifies the localization of the eigenfunctions, consistent with an enhancement of the effective confinement scale induced by the helical twist. The $m$-resolved spectra further highlight the interplay between the centrifugal contribution $\propto (m^{2}-1/4)/r^{2}$ and the torsion-induced Coulomb-like term $\propto -\,m\omega k/r$ embedded in $V_{\mathrm{eff}}(r)$, producing characteristic angular-sector splittings and distinct dispersions across $m=-1,0,1$. The corresponding probability densities confirm that the nodal structure is preserved and that the spatial support of the states is compressed as the geometric coupling strengthens.
	
	Finally, the framework developed here is readily extensible. The same Sturm--Liouville formulation and finite-difference machinery can be applied to other external interactions and to broader parameter domains (including different magnetic/flux configurations), providing a robust route to explore torsion-induced spectral engineering and confinement mechanisms in nontrivial geometries. In this sense, the present work provides both a consistent numerical baseline for helically twisted backgrounds and a practical computational template for future studies of quantum bound states in torsional/defect-like settings.

	\section{ACKNOWLEDGMENTS}

	The authors would like to thank UEMA (Universidade Estadual do Maranh\~ao) and to CNPq (Conselho Nacional de Desenvolvimento Cient\'ifico e Tecnol\'ogico - Brazil). R. L. L. Vit\'oria was supported by the EDITAL N. 102/2025-PPG/CPG/UEMA and by CNPq Project No. 150420/2025-0. E. O. Silva acknowledges the support from Conselho Nacional de Desenvolvimento Cient\'{i}fico e Tecnol\'{o}gico (CNPq) (grants 306308/2022-3), Funda\c c\~ao de Amparo \`{a} Pesquisa e ao Desenvolvimento Cient\'{i}fico e Tecnol\'{o}gico do Maranh\~ao (FAPEMA) (grants UNIVERSAL-06395/22), and Coordena\c c\~ao de Aperfei\c coamento de Pessoal de N\'{i}vel Superior (CAPES) - Brazil (Code 001).

\end{document}